\documentclass[12pt]{aastex}

\usepackage{graphics,epsfig}
\usepackage{natbib}
\usepackage{lscape}

\def\spose#1{\hbox to 0pt{#1\hss}}
\def\lta{\mathrel{\spose{\lower 3pt\hbox{$\mathchar"218$}}
     \raise 2.0pt\hbox{$\mathchar"13C$}}}
\def\gta{\mathrel{\spose{\lower 3pt\hbox{$\mathchar"218$}}
     \raise 2.0pt\hbox{$\mathchar"13E$}}}

\def\figure#1#2 {\par{\narrower\noindent {\bf Fig. #1}
   \hskip 2mm #2\par}\bigskip\noindent}
\def\table#1#2 {\par{\narrower\noindent {\bf Tab. #1}
   \hskip 2mm #2\par}\bigskip\noindent}

\def\registered{{\ooalign{\hfil\raise .00ex\hbox{\scriptsize R}\hfil\crcr\mathhexbox20D}}}

\usepackage{amsmath}
\usepackage{accents}
\newlength{\dhatheight}

\shorttitle{Exocomets in the System of HD~10180}

\shortauthors{Loibnegger et al.}

\begin{document}

%% LaTeX will automatically break titles if they run longer than
%% one line. However, you may use \\ to force a line break if
%% you desire.

\title{Case Studies of Exocomets in the System of HD~10180
}

\bigskip
\bigskip

\author{Birgit Loibnegger$^1$, Rudolf Dvorak$^1$, Manfred Cuntz$^2$}

\affil{$^1$Institute of Astronomy, University of Vienna, T\"urkenschanzstr. 17 \\
A-1180 Vienna, Austria}
\email{birgit.loibnegger@univie.ac.at; rudolf.dvorak@univie.ac.at}

\bigskip

\affil{$^2$Department of Physics, University of Texas at Arlington, \\
Arlington, TX 76019, USA}
\email{cuntz@uta.edu}

\bigskip

%%%%%%%%%%%%%%%%%%%%%%%%%%%%%%%%%%%%%%%%%%%%%%%%%%%%%%%%%%%%%%%%%%%%%%%%

\begin{abstract}
The aim of our study is to investigate the dynamics of possible comets in the
HD~10180 system. This investigation is motivated by the discovery of exocomets in various
systems, especially $\beta$ Pictoris, as well as in at least ten other systems. Detailed
theoretical studies about the formation and evolution of star--planet systems
indicate that exocomets should be quite common. Further observational results are expected in the
foreseeable future, in part due to the availability of the Large Synoptic Survey
Telescope. Nonetheless, the Solar System represents the best studied example
for comets, thus serving as a prime motivation for investigating comets in HD~10180 as well.
HD~10180 is strikingly similar to the Sun. This system
contains six confirmed planets and (at least) two additional planets subject to
final verification. In our studies, we consider comets of different inclinations and
eccentricities and find an array of different
outcomes such as encounters with planets, captures, and escapes. Comets with
relatively large eccentricities are able to enter the
inner region of the system facing early planetary encounters. Stable comets
experience long-term evolution of orbital elements, as expected. We also tried
to distinguish cometary families akin to our Solar System but no clear
distinction between possible families was found. Generally, theoretical and
observational studies of exoplanets have a large range of ramifications, involving
the origin, structure and evolution of systems as well as the proliferation of water
and prebiotic compounds to terrestrial planets, which will increase their chances
of being habitable.
\end{abstract}

\keywords{astrobiology --- circumstellar matter --- comets: general ---
          methods: numerical --- protoplanetary disks ---
          stars: individual (HD~10180)}

\clearpage

%%%%%%%%%%%%%%%%%%%%%%%%%%%%%%%%%%%%%%%%%%%%%%%%%%%%%%%%%%%%%%%%%%%%%%%%

\section{Introduction}
\label{sec:intro}

Comets appear to constitute a crucial component of many (or most) star--planet
systems as informed by the structure of the Solar System as well as observations
of the environments of stars other than the Sun.  There are also theoretical
studies, which tend to indicate that comets might be a quasi-universal
phenomenon, especially regarding the environments of solar-type main-sequence
stars (see below).  As identified in the Solar System, comets are relatively
small icy objects, often only a few kilometers in extent (i.e., $\sim$0.1
to $\sim$100~km in diameter) that formed in the outer Solar System where
temperatures have been (and still are) sufficiently low to allow for frozen water.
Comets thus represent leftovers from the early Solar System, closely associated
with its formation process \citep{bra08,coo16}.

This picture has detailed implications for Earth (and possibly Mars, if
considered in an exobiological context).  Based on detailed studies on
the early history of the Solar System, the inner part of the protostellar disk,
due to its proximity to the Sun and hence shaped by both high temperatures and
high solar wind pressure, contained a relatively high concentration of silicates
and heavy elements.  On the other hand, icy particles and water in notable
quantities could only have existed relatively far away from the Sun
\citep[e.g.,][]{mar14}.  Considering that the Earth is assumed to have formed
in the inner, dry part of the disk \citep[e.g.,][]{mar14,rub15}, most of its
water had to be delivered to it (and to Mars) from beyond the snowline, i.e.,
$\sim$3.1~au \citep{mar12}.

Nonetheless, underlying scenarios, including those considering the efficiency
of water transport by planetesimals, asteroids and comets, are still subjects
of ongoing research \citep[see][]{gen08}.  Recent observations of comets
\citep[e.g.,][]{har11,alt15} reinforce that the D/H ratio of the Solar System
comets agrees with or exceeds the D/H ratio found in Earth's oceans.  \cite{boc12}
measured, by using the {\it Herschel Space Observatory}, the $^{16}$O/$^{18}$O and
D/H ratios for the Oort Cloud comet C/2009 P1 (Garradd), whereas \cite{alt15}
focused on 67P/Churyumov-Gerasimenko, a Jupiter family comet.  Another pivotal
aspect of that latter comet is its organic-rich surface as seen by {\it VIRTIS/Rosetta}
\citep{cap15,ric15}.  Note that some of these concepts, including those based on results
from spectroscopy, are also relevant for studies beyond the Solar System.

There are also significant findings based on theoretical work. \cite{ste90}, for
instance, assumes that comets are a natural product of planetary system formation
as is the Oort Cloud. Simulations of planet formation for our own Solar System show
that during the formation process of the gas giants frequent ejection of icy
planetesimals from the planetary region took place.  Some of those led to the
formation of the Oort Cloud while most of them escaped to the interstellar space.
Based on the Nice model of planet migration, \cite{bra08} investigated the possible
two-stage formation of the Oort Cloud in the Solar System.  Their model showed a
possible mass for the outer Oort Cloud in the range of 0.5 to to 1 $M_\oplus$ with
a most likely value close to 0.9 $M_\oplus$, in agreement with observational
estimates.  This finding is also relevant for the possibility of comets in
exosolar systems.

In fact, it is noteworthy that pivotal evidence has been obtained that comets
exist around stars other than the Sun as well; they are also referred to as
exocomets or Falling Evaporating Bodies (a somewhat more generic term).
The first system named to harbor exocomets is $\beta$~Pictoris, an A6 main-sequence
star.  \cite{beu90} were first to analyze the metallic absorption lines found in the
stellar spectrum and to create a model of evaporation, which fitted the behavior
of the Ca~II.  They argued that frequent infalls could be the result of a
perturbing action of an (at that time) protoplanetary body embedded in the disk.
Furthermore, they related the dynamics of the exocomets-as-indicated to a major
planet in the $\beta$ Pictoris system, which was not known at the time, but
has been found later ($\beta$~Pictoris~b).

A larger array of observational results regarding comets in the $\beta$~Pictoris
system has recently been given by \cite{kie14}.  This work made use of more than
1,100 spectra (Ca~II~H,~K) obtained between 2003 and 2011 by the HARPS ({\it High
Accuracy Radial velocity Planet Searcher}) instrument.  In fact, as described
by the authors, the  $\beta$~Pictoris system appears to contain two families
of exocomets, i.e., one consisting of exocomets producing shallow absorption lines,
attributable to old exhausted comets trapped in a mean motion resonance with a
massive planet (as previously indicated by \cite{beu90}), and one family of
relatively recent comets associated with the fragmentation of one or a few
parent bodies.
Based on results by, e.g., \cite{zuc12} and \cite{wel13}, comets have also
been found in other exoplanetary systems as well.  One example is 49~Ceti, a
relatively young (40~Myr) A-type main-sequence star.  Additional examples are
HD~21620, HD~42111, HD~110411, HD~145964, and HD~183324 (among others), which
increases the total number of stars with indications of exocomets to (at least) 10.
Several of these cases have been identified by \cite{wel13} who studied Ca~II~K
absorption profiles from 21 nearby A~stars with circumstellar gas debris disks.
They found weak absorption features, which appear only sporadically and show
radial velocities in the range of ${\pm}100$~km~s$^{-1}$, which the authors
interpret as firm indications of exocomets.  The indicated exocomets are
around mainly young stars ($\sim$5~Myr); these stars are supposedly in the
phase of planet formation.

There are also significant astrobiological implications to the study of exocomets,
which are closely related to the possible habitability of terrestrial planets in
exosolar systems.  A highly significant aspect is that exocomets are able to deliver
both liquid water and basic organic substances from regions beyond the snowline to
the inner domains of star-planet systems where terrestrial planets with features
in supportive to habitability reside; see, e.g., observational results and analyses
for 67P/Churyumov-Gerasimenko by \cite{cap15} and \cite{ric15} based on {\it Rosetta}. 
Recently, \cite{bos16} argued that the ability of comets and comet-like asteroids to foster
prebiotic chemistry as well as their available energy may be relevant for
facilitating habitability in various exosolar systems.

A more stringent perspective has previously been conveyed by \cite{wic09} and \cite{wic10}.
They discussed the possible existence of liquid water in comets, including the
general potential for panspermia \citep[e.g.,][]{hoy86,blo03}.  Based on the
temperature and radiative environments, transient melting in comets in inner
stellar systems might be able to occur.  Supposed that comets were seeded with
microbes at the time of their formation from prebiotic material, \cite{wic09}
conjectured that there would by sufficient time available for exponential
amplification and evolution within the liquid interiors.  From a terrestrial
point of view, certain kinds of extremophiles have been identified
\citep[e.g.,][]{rot01}, which could potentially exist and flourish
in such extreme environments.

The general aspect of our work is to explore based on theoretical orbital simulations
whether in exosolar systems as, e.g., HD~10180 (where the center star has properties
akin to the Sun) comets happen to exist in different families similar to the ones
observed in the Solar System, namely, the Halley comets (HC) and the Jupiter family
comets (JFC). However, it turned out that for HD~10180 no clear
separation of cometary families could be identified. In Section~2, we discuss our
theoretical approach, including comments about the HD~10180 system itself with focus
on the Neptune-type planets considered for our time-dependent simulation.  We also
describe the methods as used and the numerical set-up.  Our results and discussion
are given in Section~3.  Finally, in Section~4, we present the summary and conclusions
of our work.

%%%%%%%%%%%%%%%%%%%%%%%%%%%%%%%%%%%%%%%%%%%%%%%%%%%%%%%%%%%%%%%%%%%%%%%%%%%%%%

\section{Theoretical Approach}
\label{sec:theo}

\subsection{The HD~10180 System}
\label{sec:system}

HD~10180 is located in the southern-sky constellation of Hydrus at a distance
of $39.0 \pm 0.6$~pc from the Sun \citep{lee07}.  Based on the {\it Hipparcos}
catalog \citep{esa97}, the stellar spectral type is given as G1~V \citep{nor04},
which makes HD~10180 an almost-twin to our Sun (G2~V).  The stellar effective
temperature is given as $5911 \pm 19$~K, and its luminosity and mass are
identified as $1.49 \pm 0.02$~$L_\odot$ and $1.06 \pm 0.05$~$M_\odot$,
respectively; see \cite{lov11} and references therein.  Based on the star's
level of chromospheric activity, the age of HD~10180 was determined as
$4.3 \pm 0.5$~Gyr \citep{lov11}, which is (within its error bar) in agreement
with the age of the Sun.

\cite{lov11} conducted a high-precision radial velocity survey with the HARPS
spectrograph, which led to the discovery of a planetary system consisting of at
least five Neptune-mass planets with minimum masses ranging from 12 to 25~$m_\oplus$
(i.e., planet c, d, e, f, and g).
They also found tentative evidence for an inner Earth-mass planet (HD~10180~b)
with a minimum mass of $1.35 \pm 0.23~M_{\oplus}$ as well as more substantial
evidence for an outer, more massive planet (HD~10180~h) with a minimum mass of
65~$m_\oplus$ located at a distance of $3.42^{+0.12}_{-0.13}$~au, and with an
approximate orbital period of 2248~days; see Table~\ref{Table1} for details.
In a succeeding study, \cite{tuo12} found two statistically significant
signals corresponding to two additional planets in close circular orbits
(i.e., HD~10180~i and HD~10180~j) with $67.55^{+0.68}_{-0.88}$
and $9.655^{+0.022}_{-0.072}$ day periods and minimum masses of
$5.1^{+3.1}_{-3.2}~m_{\oplus}$ and $1.9 ^{+1.6}_{-1.8}~m_{\oplus}$,
respectively, indicatives of super-Earths.

\cite{tuo12} also confirmed the existence of the five Neptunian planets
(with minor revisions concerning the previously reported parameters) as well as
the existence of HD~10180~h and found evidence for two additional planets.
Thus, based on the work by \cite{lov11} and \cite{tuo12}, it can be concluded
that HD~10180 is a tightly packed system with six or possibly nine planets;
eight of those are spaced between 0.02 and 1.42~au from the star and another
planet is located at $\sim$3.40~au.  If all of these planets are verified,
it would make HD~10180 the present-time record holder with more planets in
orbit than there are in the Solar System!  Note that the semi-major axes are
fairly regularly spaced on a logarithmic scale, thus exhibiting an approximate
{Titius-Bode}--type law.  Other examples of planetary systems exhibiting
quasi-logarithmic spacing have previously been explored by, e.g., \cite{cun12}
and \cite{bov13}.  \cite{lov11} also commented on possible formation scenarios
of the HD~10180 system in the view of the planetary spacing and mass distribution.

Reasons for us to choose HD~10180 as system of study include the striking
similarity of the host star to the Sun and, even more importantly, the system's
general composition:  The most massive planet HD~10180~h is located outside
of all other planets at a distance of $\sim$3.4~au from the central star with
a mass of $\sim$64.4~$m_{\oplus}$; all inner planets, including the five
Neptunian planets have masses $m < 26~m_{\oplus}$ and four of them orbit
inside of 1~au.  Although there are notable differences to the Solar System,
there are general properties where the system of HD~10180 and that
of the Sun agree:  in the Solar System, Jupiter, the most massive planet
(317.8 $m_{\oplus}$), acts as the main perturber for asteroidal and
cometary objects, thus deflecting them to reach the terrestrial planets
located inside of its orbit.  Therefore, in the HD~10180 system, we expect
HD~10180~h to play a similar role resulting in deflecting and capturing comets.

In Table~\ref{Table2}, further information is given about the four outer
planets HD~10180~e to h considered in our computations.  The Hill radius of
HD~10180~h extends to $20 \times 10^6$~km (i.e., 53 times the Earth--Moon
distance).  Jupiter in our Solar System, with a semi-major axis of 5.2~au
and about 4 times the mass of HD~10180~h has a Hill radius $R_{\rm Hill}$
$\simeq 52 \times 10^6$~km.  In the following, two different densities are
assumed for the planets of HD~10180, which are: $\rho_1 = 1$~g~cm$^{-3}$
and $\rho_2 = 2$~g~cm$^{-3}$; the assumed density values are relevant for
investigating close comet--planet encounters.  Here $\rho_1$ is set to be
very close to the numerical average between the densities\footnote{
For comparison: Jupiter: 1.326~g~cm$^{-3}$, Saturn: 0.687~g~cm$^{-3}$, and
Neptune: 1.638~g~cm$^{-3}$} of Jupiter and Saturn, whereas $\rho_2$ is
chosen as the approximate upper limit of medium-sized gas planets
as, e.g., Neptune.  The choices for $\rho_1$ and $\rho_2$ are also
consistent with empirically deduced values for gaseous exoplanets obtained
by \cite{gui99} and subsequent work.

\subsection{Methods and Numerical Setup}
\label{sec:methods}

Guided by the structure of the Solar System, we assume that the system of
HD~10180 might also harbor an Oort-type cloud of comets far from the star,
and perhaps other comets in closer stellar vicinity (i.e., comets up to 1000~au
from the star and Kuiper belt comets).  It is further assumed that every now
and then the comets are gravitationally disturbed by, e.g., a passing star
or a cloud of interstellar matter.  The focus of our study is to investigate
how those comets are disturbed or captured by one of the close-in planets,
or even forced to escape from the system.
When referring in our study to comets as captured, we address comets
that are for a certain timespan in an orbit of small eccentricity and
semi-major axis.

Our computations take into account a total of four planets (see Table~\ref{Table1}),
considering that the inner planets (with the closest planet having a semi-major
axis of only $a \simeq 0.06$~au) are not expected to noticeably change our
statistical results. A more comprehensive approach based on additional planets would be without an obvious impact on the outcome, as our test computations have shown. It is evident that the outermost planet (i.e., HD~10180~h), considered in
our simulations, is expected to be the dominant planet affecting the
exocomets' orbital motions, a role played by Jupiter in the Solar System.

In our study, we used extensive numerical integrations for tens of thousands
of fictitious comets (assumed as massless) for the HD~10180 system.  The method
of integration for the equations of motion is based on Lie-transformation, which
is known to be a fast and efficient tool.  This method has been successfully used
since many years for different problems in astrodynamics \citep[e.g.,][]{dvo86}.
Comparisons with other methods have shown that this method is especially
adapted to model close encounters of celestial bodies as well as collisions.
The reason is that this method utilizes an automatic step-size control
connected to the chosen precision selected for solving the differential
equations \citep[e.g.,][]{egg10}; it was also compared to other methods
in numerous ways.

Our integration scheme encompasses the central star of the HD~10180 system,
the four system's outer planets (see Table~\ref{Table1}) as well as 100 fictitious
exocomets at an initial distance of $a=89$~au, assumed as massless. 
The comets were assumed to originate from an Oort Cloud analogue and
therefore have been placed in nearly hyperbolic orbits.  The initial semi-major
axis of 89~au was chosen to be close enough to allow interactions with the system's
inner planets, but, on the other hand, also set to be well beyond the outermost
system planet, HD~10180~h.  Different initial semi-major axes were used in test
simulations for some of the initial conditions.  However, no discernible differences
in the final statistics for the comets were found.

The initial inclinations and eccentricities
are chosen as follows: initial inclinations from the plane ($i =1^{\circ}$)
with ${\Delta}i = 10^{\circ}$ extending till retrograde orbits for 100 different
perihelion longitudes and initial eccentricities ranging from $e_0=0.905$ to
0.99. Furthermore, the longitude of the ascending node $\Omega$
was set to the same angle as the inclination and the mean anomaly $M$ was always
set to $1^{\circ}$. With these choices of initial conditions the number of integrated
fictitious cometary orbits was more than 30,000, namely 19 ($i$) $\times$
100 ($\omega$) $\times$ 16 ($e$) and we were able to cover all the possible directions
of incoming comets (see Fig.~\ref{Fig2}). The integration time was set to 1~Myr;
in a few cases of special interest, we continue to 5~Myr (see Fig.~\ref{Fig3}
and~\ref{Fig4}). 

\begin{figure*} 
\centering
\begin{tabular}{c}
\epsfig{file=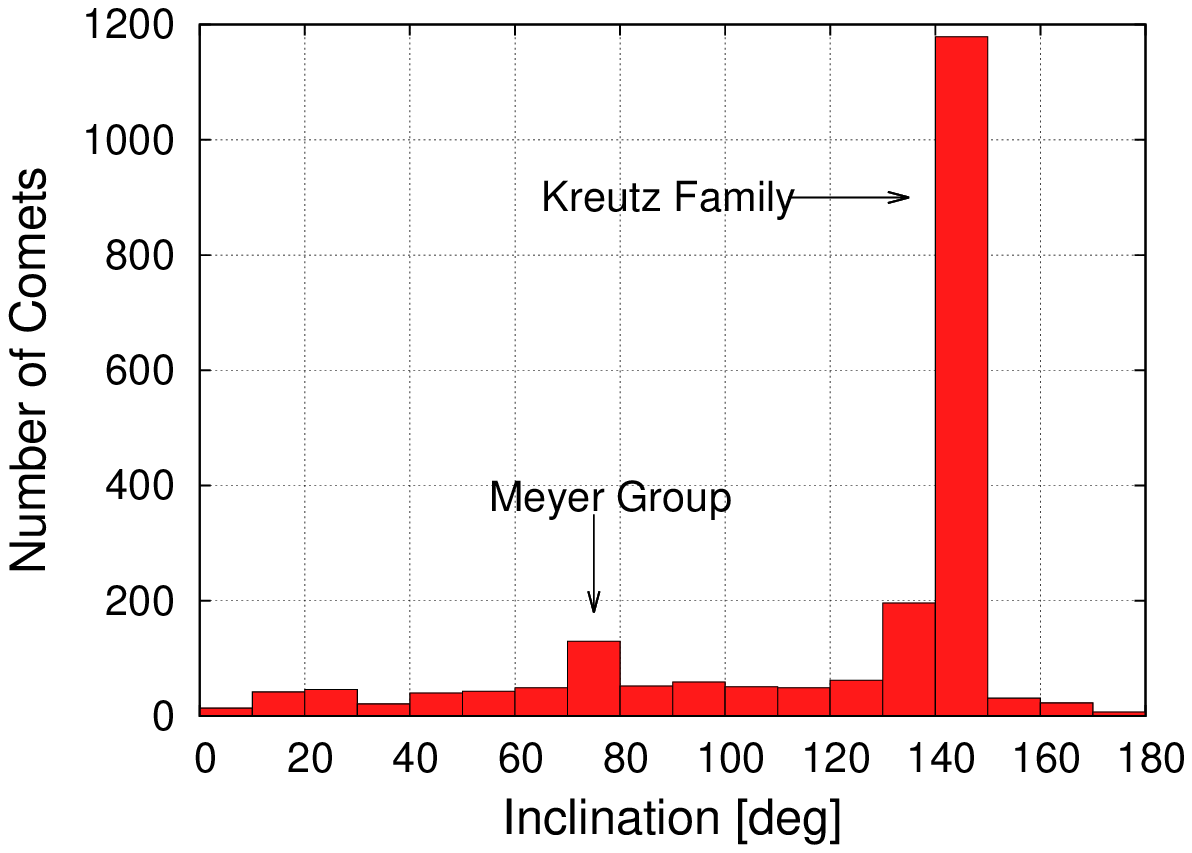,height=10cm,width=13cm}
\end{tabular}
\caption{
Distribution of comets in the Solar System with respect to their inclination
on their hyperbolic or nearly parabolic orbits. The Kreutz family of comets
is clearly visible as they all show an inclination around $145^{\circ}$.
Except for this peak, all other inclinations show roughly the same numbers.
Motivated by this result, our initial conditions for HD~10180 are also based
on a spherical distribution of comets. See also
{\tt Jssd.jpl.nasa.gov/sbdb-query.cgi} for more information.
}
\label{Fig1}
\end{figure*}

\begin{figure*} 
\centering
\begin{tabular}{c}
\epsfig{file=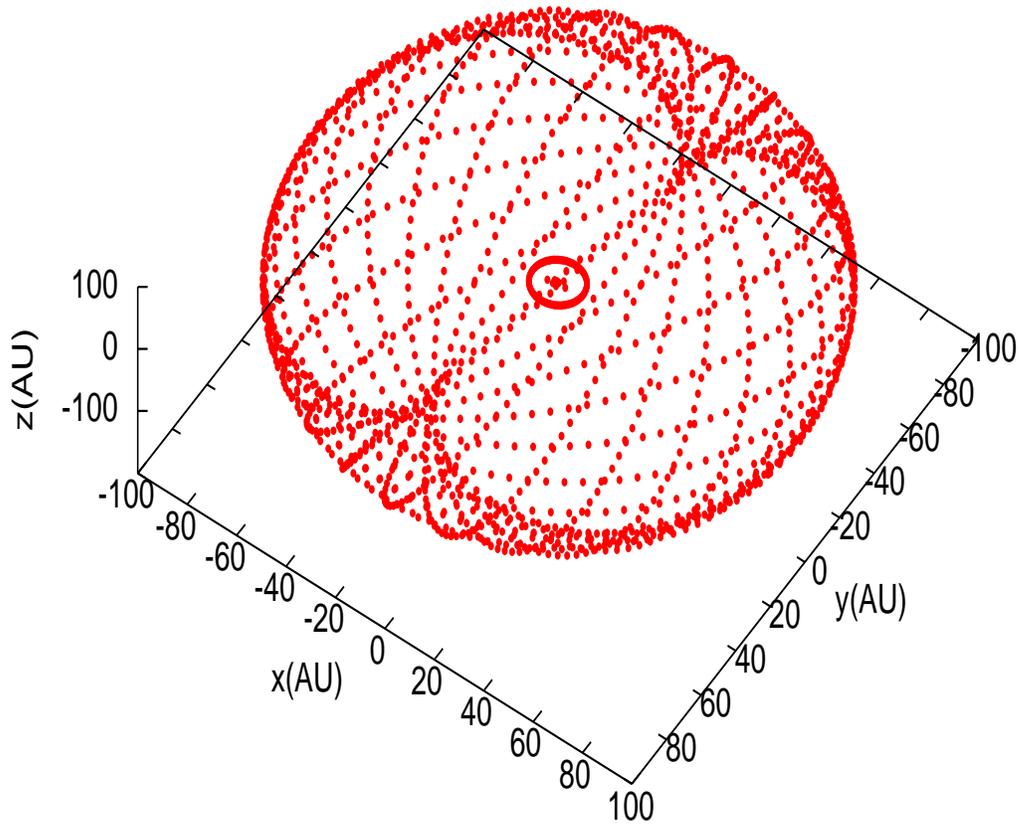,height=15cm,width=12cm,angle=270}
\end{tabular}
\caption{
Initial distribution of comets from $i_0=0^\circ$ to $180^\circ$ and $a=89$~au
in our integrations; the narrow ring close to the central star indicates
the orbit of planet HD~10180~h.
}
\label{Fig2}
\end{figure*}

It could be argued that in our Solar System many of the comets have orbits of
low inclination --- so why consider an isotropic influx of comets in HD~10180?
The reason is that without detailed information of possible exocomets in that system,
an isotropic angular distribution for the initial conditions appears to be the most
appropriate default assumption.  Figure~\ref{Fig1} shows the distribution
of comets in the inner Solar System on hyperbolic or nearly parabolic orbits,
which can then be readily observed. The results from the SOHO satellite are
indicating that there are two cometary families: the Kreutz sungrazers
as well as the Meyer group of comets.  Other comets entering the Solar System
seem to be randomly distributed over the sphere.  As part of our study, we
wish to identify the prevailing patterns of comets for HD~10180.

During orbital evolution, the orbits of the exocomets undergo close encounters
with the planets.  This behavior sometimes led to captures into low-eccentric
orbits (see Fig.~\ref{Fig3} and~\ref{Fig4}) or to ejections from the system;
in a few cases, collisions with the host star occurred.  Whenever a comet was
ejected from the system, we inserted another one with the same `initial'
conditions as the escaper.  Since at the time of insertion the configuration
of the planets has changed, the newly inserted comet will exhibit a different
dynamics.  Based on this procedure, we will always keep the number of fictitious
exocomets equal to 100.

As part of our numerical simulation, we counted the unstable comets for the various 
eccentricities and inclinations; the respective values are shown in Fig.~\ref{Fig5}.
If all comets stayed on stable orbits --- they could still be captured as periodic
comets (e.g., Fig.~\ref{Fig7}) or assume a chaotic but still stable orbit --- the
number of ejected comets in the respective table is zero.  Figure~\ref{Fig5} shows
very well the regime of stable orbits for comets of initial inclinations $100^{\circ}< i < 160^{\circ}$ and eccentricities $e_0 < 0.99$ surrounded by areas of initial conditions where many orbitally unstable comets found.

%%%%%%%%%%%%%%%%%%%%%%%%%%%%%%%%%%%%%%%%%%%%%%%%%%%%%%%%%%%%%%%%%%%%%%%%%%%%%%

\section{Results and Discussion}
\label{sec:results}

\subsection{General Case Studies}
\label{sec:studies}

By making use of Lie-series
\citep[see, e.g.,][for background information]{gro60,stu74,dvo83,del84,egg10}
close encounters of the incoming comets with the planets are calculated with a
sufficiently high precision.  Due to close planetary encounters, some of the
comets are captured, whereas others are ejected from the system.  In the
Solar System, captured comets form families due to the gravitational influence
of Jupiter. Our simulations show that the outermost planet (i.e., HD~10180~h)
is able to change the orbits of incoming comets most profoundly when arriving
with high speeds from the outskirts of the system.  Moreover, close encounters
also occur with respect to the inner planets as considered (see Table~\ref{Table1}).

The number of ejected comets increases more significantly for comets of
larger eccentricities and smaller inclinations, owing to the fact that (1) for
smaller inclinations, the number of encounters and consequently escapers is
larger and (2) with larger eccentricities, there is a an increased likelihood
that the comets enter the inner region of the system and undergo relatively
early encounters resulting in escape.

The combined preliminary results for inclinations from $i_0=0^{\circ}$ to $90^{\circ}$
are shown in Tables~\ref{Table3} and~\ref{Table4} (prograde orbits), as well as in
Table~\ref{Table5} from $i_0=100^{\circ}$ to $170^{\circ}$ (retrograde orbits) and
in Fig.~\ref{Fig5}.  It is found that for large eccentricities many more escapes
from the system occur.  The border between completely stable and escaping orbits
is not a sharp one; in fact, at the interface separating the two regions, there
is a small band characterized by statistical number fluctuations.  However, there
are no escapers for retrograde orbits up to $e_0=0.94$; hence, we don't show this
regime in a separate table.  Also, this `mixed' band is still under investigation,
as well as the computations for $i_0=180^{\circ}$.  But even when there is no immediate
crossing of the cometary orbit with the planet, the impact of secular resonances can
lead to changes in a comet's orbit that close encounters with the planets are possible.
Figure~\ref{Fig6} reveals that the orbit for $e_0=0.935$ (dark blue line) is in terms
of its perihelion distance still far away from the orbit of HD~10180~h ($>2$~au),
but nevertheless at a later time, this orbit is changed so that close encounters
with the planets lead to the comet's escape from the system.  Table~\ref{Table3}
shows that a total of 104 comets have been ejected from the system (first line for
$i_0=0^\circ$ and $e_0=0.935$).

In Fig.~\ref{Fig11} to \ref{Fig13} we show orbital properties of comets
populating orbits of capture, depending on their initial conditions.
Only for a few cases it is possible that comets
of initially planar orbits are also spread to retrograde orbits as the simulations
progress.  The different values of the initial eccentricities depicted in the
figures are $e_0=0.99$, 0.98, 0.975, and 0.965.  Together with the chosen semi-major
axis of $a=89$~au, these values make those comets immediately to possible planet
HD~10180~h ($a=3.4$~au) crossers.  The encounters with the planets force the comets
into either ejected orbits or captured ones. 

%Reformulated: after referee advice:

%1)
The effect of the most massive planet, HD~10180~h (see Table~\ref{Table1}),
on comet scattering seems obvious. Nonetheless, the second massive planet, HD~10180~g,
located at about half the distance from the star, could potentially also have a
non-negligible influence on cometary orbits.  This can be seen in the graphs of
Fig.~\ref{Fig8}, which show the main outcome of our simulations. In the
{\it lower left panel} of Fig.~\ref{Fig8} (integration done without planet
HD~10180~h) and Fig.~\ref{Fig8} {\it upper right panel} (integration done without
planet HD~10180~g) the differences are well pronounced. When planet h is excluded,
only the comets with initial eccentricity of $e > 0.98$ are able to reach the
orbit of planet g and thus get scattered and ejected from the system. The influence
of planet g, on the other hand, is not that big. Comparing the two {\it upper} panels
of Fig.~\ref{Fig8} shows that they look almost identical, which means that taking out
planet g does not have a noticeable effect on the scattering process.  Nevertheless,
there is still a difference in the maximal number of ejected comets. Taking out
planet g reduces the total number of ejected comets. Integrating the system for
5 Myr (shown in Fig.~\ref{Fig8} in the {\it lower right} panel) demonstrates that
only the total number of ejected comets increases. However, the overall picture of
stability does not noticeably change from a statistical point of view.  Comets with
initially high eccentricity and low inclination $e_0 \, > \, 0.5$ and
$0^{\circ} \, < \, i_0 \, < \, 40^{\circ}$ tend again to
become unstable, whereas the main regime of initial conditions $(e_0,i_0)$ leads
to stable cometary orbits.  These comets stay in the system for the entire integration
time of 5~Myr.

%2)

Figures.~\ref{Fig11} to \ref{Fig13} show the properties of captured comets
based on 1~Myr computations by depicting various orbital elements, which are:
eccentricity, semi-major axis, and inclination (see color code).  While for
Fig.~\ref{Fig11} all 4 planets were included, planet g was excluded from the
simulations depicted in Fig.~\ref{Fig12} and planet h was excluded for the
simulations of Fig.~\ref{Fig13}.  It is found that the difference in the scattering
outcome for Fig.~\ref{Fig11} and Fig.~\ref{Fig12} is relatively minor.  Comets with
eccentricities as shown are able to reach the inner parts of the planetary system and
interact with the planets; in fact, they are scattered to orbits with eccentricities
of $e < 0.6$ and semi-major axis as small as 5~au. Interestingly, the inclinations
of the final orbits of comets can remain high, especially for comets with initially
high eccentricities.  This behavior is revealed by the orange color of the dots in
the {\it lower right panel} of Figs.~\ref{Fig11} and \ref{Fig12}.

Fig.~\ref{Fig13}, however, indicates a completely different outcome.  For the
integrations as depicted, planet h has been excluded from the computations.  It can be
seen that comets of relatively low initial eccentricity ($e_0=$0.965) are unable to
reach the orbit of planet g, which is now the outermost planet able to influence the
orbits of incoming comets.  Thus, the comets stay on their highly eccentric orbits far
away from the planetary system.  Increasing the initial eccentricity shows that comets
can now approach planet g and can be scattered to orbits farther inward, i.e.,
$a < 60\,$~au.  Note that only comets with an eccentricity as high as 0.99 can be
scattered to orbits with $e < 0.6$ and $a < 5\,$~au.  Nevertheless, the number of
comets ending up in such orbits is small compared to the number of comets based on
the integration for the whole system versus the system with planet g excluded.
This leads again to the conclusion that the influence of the most massive planet h
on comet scattering is clearly dominant.

%3)

Furthermore, when considering the interval of $0.1 < e < 0.7$ in Figs. \ref{Fig11} and
\ref{Fig13}, one can see that with planet h included, more comets are scattered into
orbits with an eccentricity of that range.  However, with planet h excluded,
the comets are not able to reach a planetary orbit, which would influence them
gravitationally and scatter them to orbits of smaller eccentricity
(see Fig.~\ref{Fig13}).  The significant influence of planet HD~10180~h on the
scattering of the comets, especially on comets of higher initial inclination, can be
explained by the size of its Hill sphere.  Due to its bigger mass and its orbit, the
Hill sphere of planet HD~10180~h has a radius of 0.1379~au, which is more than
three times larger than that of planet g, given as 0.04~au (see Table~\ref{Table2}).

Moreover, comets interacting gravitationally with planet h have a lower
velocity, which results in a longer period during which planet h is able to affect
the comet's orbit.  Comets encountering planet g are closer to their perihelion and
thus have a higher velocity, which further reduces the duration of stay in the sphere
of influence of planet g. Thus, taking out planet HD~10180~h has a profound influence on the
scattering process. 
Additionally, it is found that the influence of planet g on the comets is relatively
weak because of its position compared to planet h.

%4)

We also explored the probability of comets to be captured into orbits
with moderate values for the semi-major axis and eccentricity, i.e.,
$e \,<\, 0.7\,$ and $a\, <\, 10\,$~au. This was done for two cases: first,
with all four planets included in the calculations (see Fig.~\ref{Fig14}) and
second, for a system with planet HD~10180~h excluded (see Fig.~\ref{Fig15}).
It was found that the highest probability of capture occurs for comets with
an initial eccentricity between ${e_0}=$~0.985 and 0.97 combined with an
initial inclination close to $0^{\circ}$.  The next highest peak, albeit
significantly lower, was found for comets with an initial inclination of
$30^{\circ}$.  Furthermore, for comets with initial eccentricities below
0.97, no significant captures occurred. Additionally, it was  
found that the overall picture did not change much when excluding planet h,
except that for comets with an initial eccentricity of 0.985 the small peaks
showing a capture probability of $\sim \, 5 \% \, $ for comets with an
initial inclination of $30^{\circ}$, $90^{\circ}$, and $120^{\circ}$ disappeared.
Hence, we conclude that planet g is responsible for the capture of comets with
this particular initial condition.

\subsection{A Very Special Capture}
\label{sec:special}

\begin{figure*} 
\centering
\begin{tabular}{c}
\epsfig{file=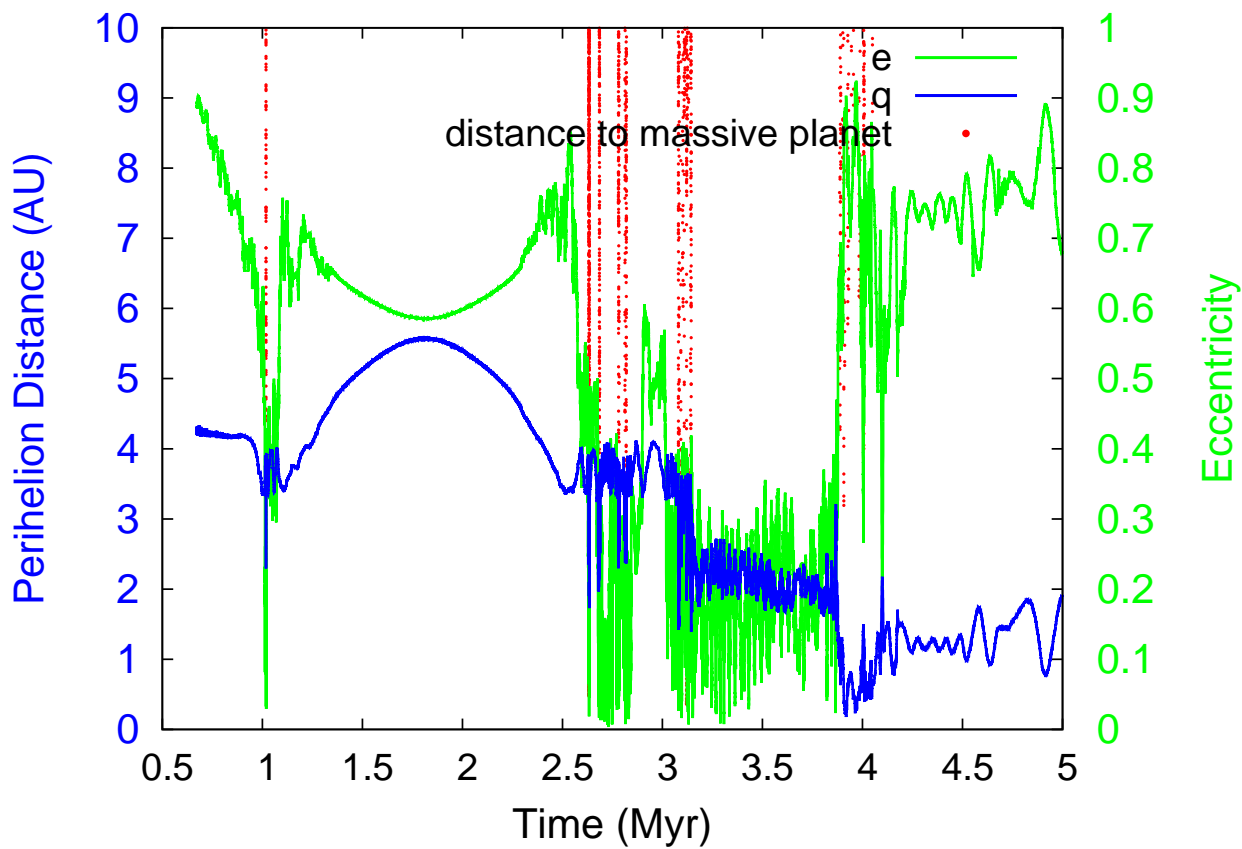,width=1.0\textwidth}
\end{tabular}
\caption{
Time evolution for a particular example of a captured comet in the
HD~10180 system.  The $y$-axis shows the perihelion distance (blue),
the eccentricity (green), and the close encounters with planet h (red)
in units of its Hill radius (see Table~\ref{Table2}).  The stable period
between 1 and 2.5~Myr is well pronounced.  After this period, another
time interval exhibiting low eccentricities follows.  These periods
are bounded by close encounters with the most massive planet in the
system, planet h.  Gravitational interactions during these encounters
lead to visible changes in the cometary orbit.
}
\label{Fig3}
\end{figure*}

\begin{figure*} 
\centering
\begin{tabular}{c}
\epsfig{file=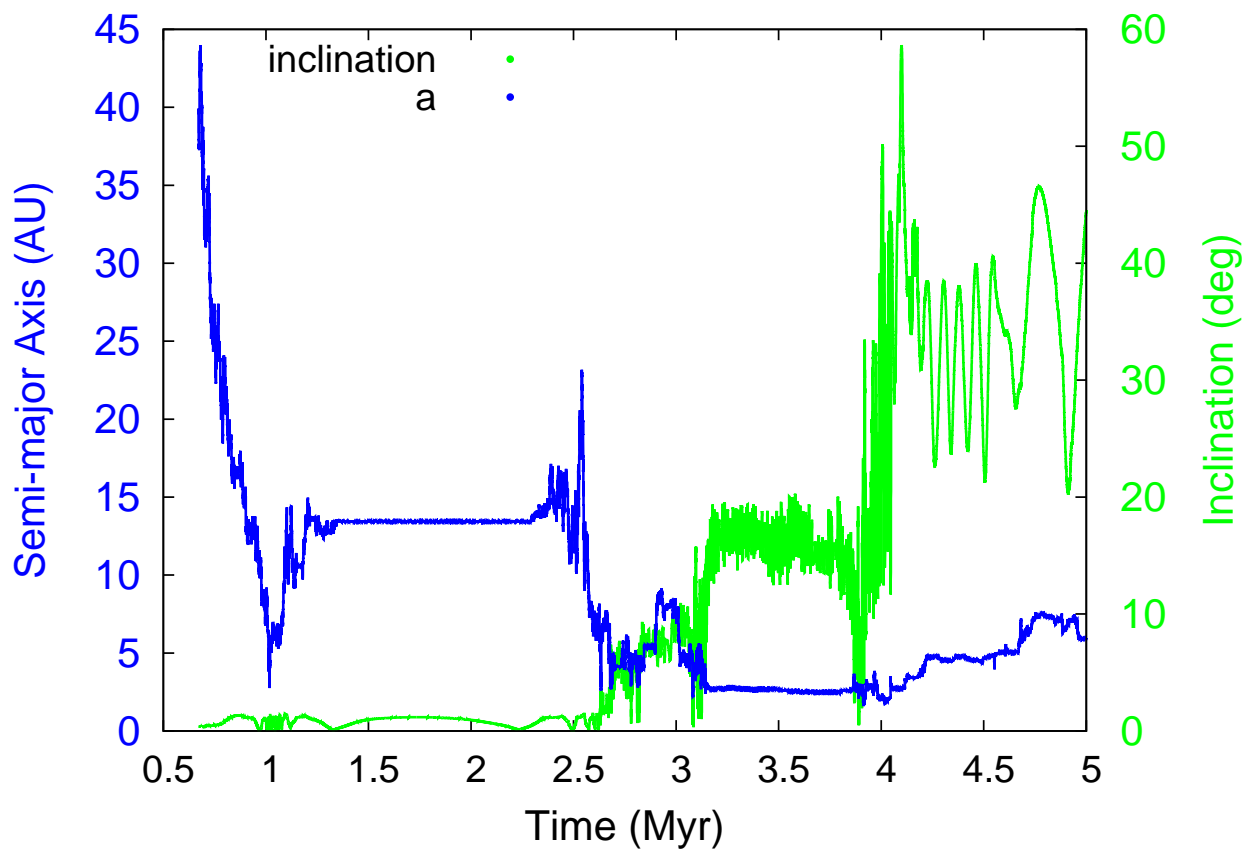,width=1.0\textwidth}
\end{tabular}
\caption{
Time evolution for a particular example of a captured comet in the
HD~10180 system.  The $y$-axis shows the semi-major axis (blue) and
the inclination (green).  The comet enters the inner region and is
captured in a stable orbit with $a \sim 14$~au for approximately 1.5~Myr
(see also Fig.~\ref{Fig3}).  Another stable period, even closer to the star
(${\sim}4$~au) follows.  After a close encounter (see Fig.~\ref{Fig3}),
the orbit becomes chaotic again and the comet may leave the inner
planetary system once more.
}
\label{Fig4}
\end{figure*}

\begin{figure*} 
\centering
\epsfig{file=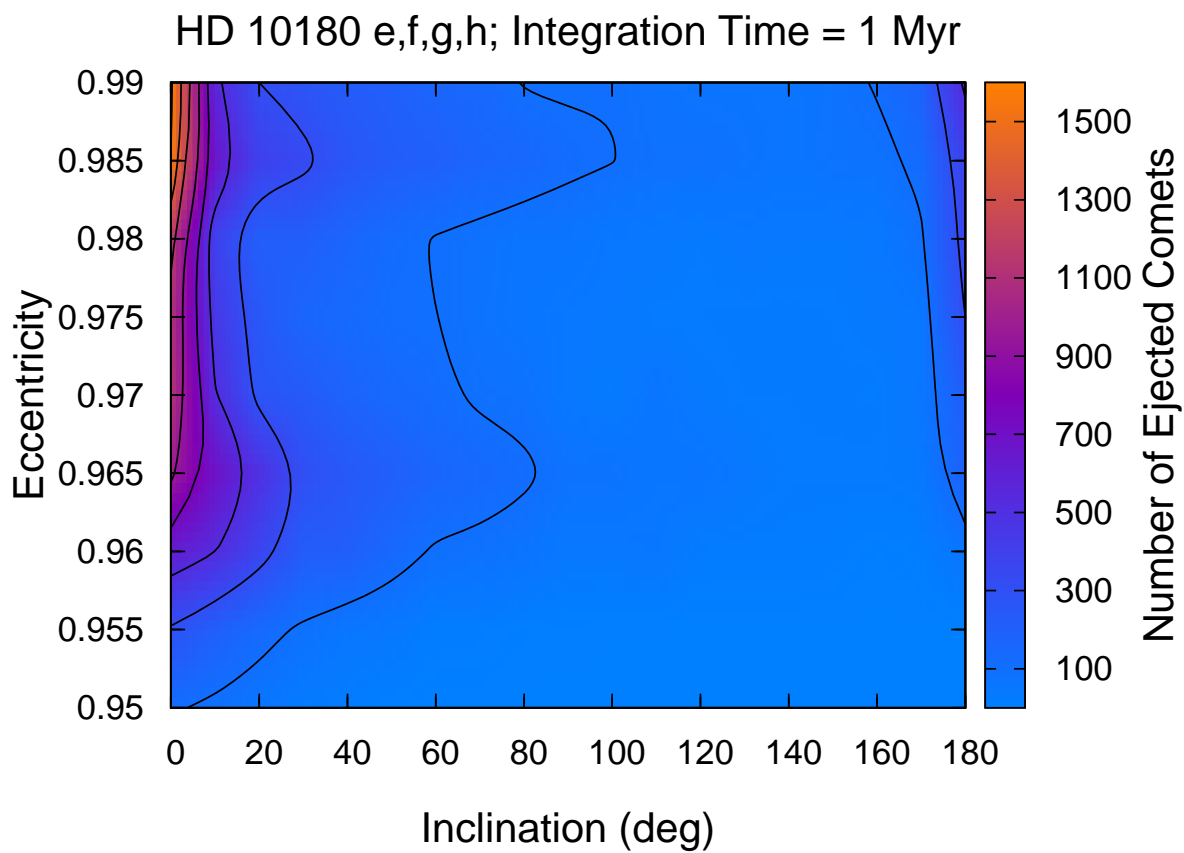,width=1.0\textwidth}
\caption{
The color scheme indicates the number of ejected comets from the system
for different initial conditions $(e_0,i_0)$ and an integration time
of 1~Myr.  It is evident that for high initial eccentricities $e_0$ and
low initial inclinations $i_0$ the number of ejected comets is highest.
Furthermore, for higher inclinations the number of ejected comets ---
even those with high eccentricities --- decreases.  This means that comets
starting with initial conditions in this regime tend to be stable for the
entire integration time of 1~Myr.
}
\label{Fig5}
\end{figure*}

\begin{figure*}
\centering
\begin{tabular}{c}
\epsfig{file=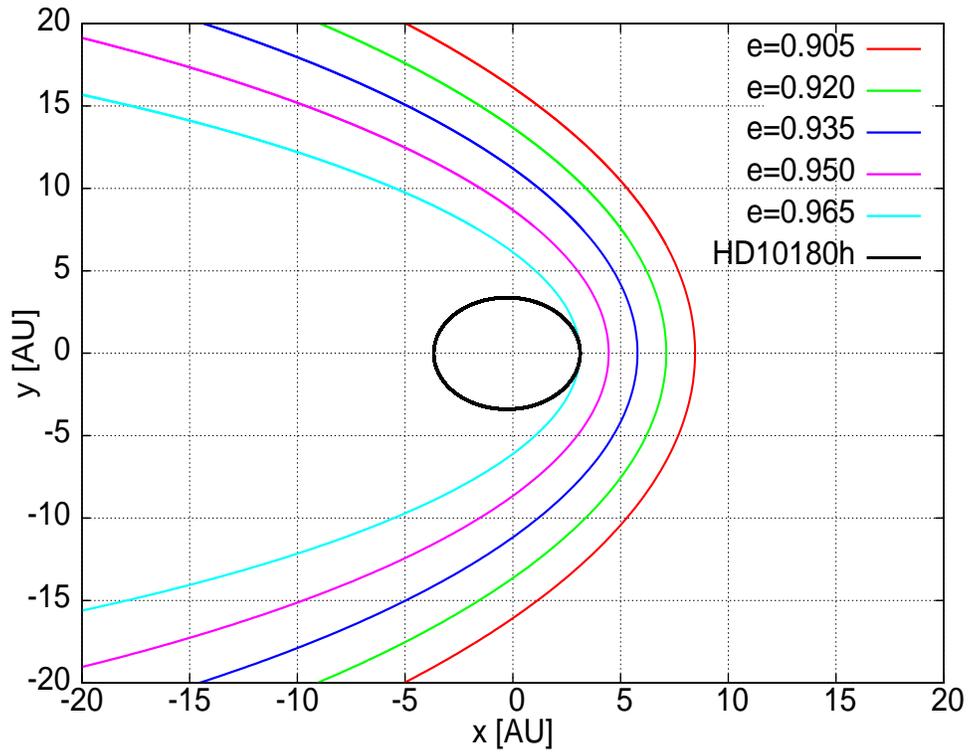,height=13cm,width=10cm,angle=270}
\end{tabular}
\caption{Depiction of five ``safe" orbits for comets due to their initial
conditions; the orbit of HD~10180~h is shown in black.  Even with a high eccentricity
of 0.965 (where the comet comes as close to the star as HD~10180~h), stable orbits
are possible.  However, as depicted in Fig.~\ref{Fig5} --- even with $e_0 \, > \, 0.95 \, $,
indicating that the perihelion distance of such a comet is still far away from
the orbit of planet h --- these comets are ejected.
}
\label{Fig6}
\end{figure*}

\begin{figure*}
\centering
\begin{tabular}{c}
\epsfig{file=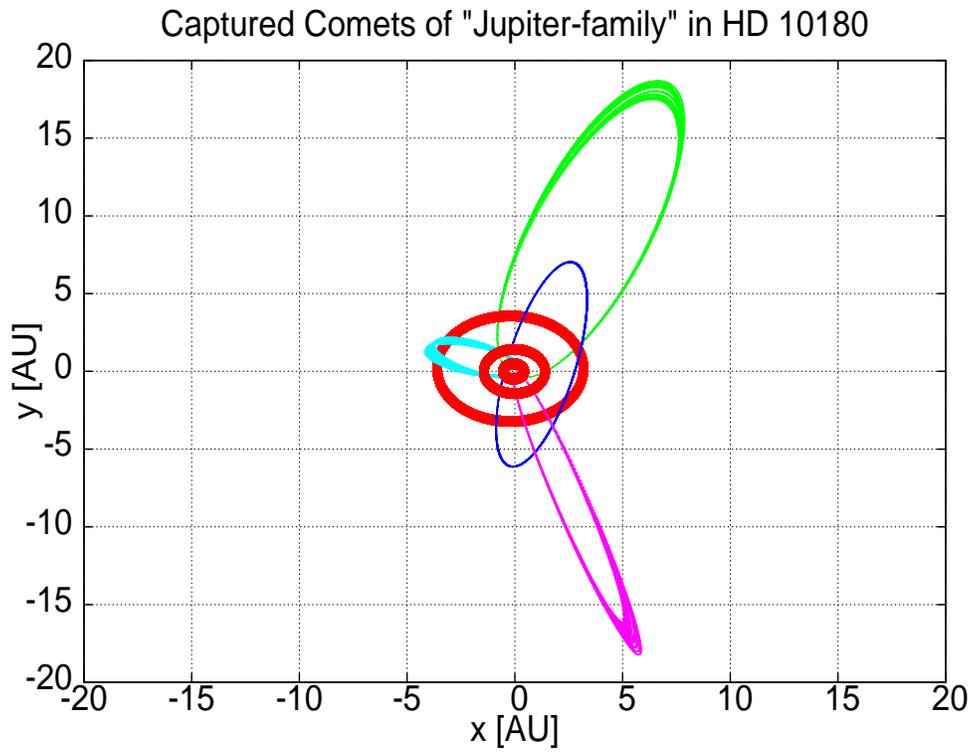,height=13cm,width=10cm,angle=270}
\end{tabular}
\caption{Orbits of four captured comets (in green, dark blue, light blue,
and purple) together with the three outer planets HD~10180~f, g, and h
(in red).
}
\label{Fig7}
\end{figure*}

Next we discuss an example of a captured exocomet in more detail (see Fig.~\ref{Fig3}
and~\ref{Fig4}).  Although the ordinary integration time has been set to 1~Myr,
we have chosen an extended time of up to 5~Myr for orbits exhibiting features of
particular interest.  This example has also been chosen because of a secular resonance,
which is still under further investigation; see \cite{lho16}. 

It is found that the comet's semi-major axis decreases drastically within
several 10$^5$~yrs from $a = 45$~au to $a = 5$~au (see {\it blue} line in Fig.~\ref{Fig4}).
Additionally, within the same time interval ($\tau < 1$~Myr) the comet's
eccentricity drops from almost parabolic ($e \simeq 1.0$) to close to circular
($e \simeq 0.0$) ({\it green} in Fig.~\ref{Fig3}). 
Thereafter, the comet's semi-major axis increases within a
relatively short time frame to up to about $a=15$~au, followed by some fluctuations
between 9 and 15~au ({\it blue} in Fig.~\ref{Fig4}).
The decrease of the comet's semi-major axis (shown in {\it blue} in Fig.~\ref{Fig4})
can be explained by jumping from one resonance to another.  Note that the same behavior
was previously observed for the Solar System comet Halley \citep{dvo90}.

Subsequently, the semi-major axis
settles at $a = 13$~au due to a close encounter with planet HD~10180~b at $1.3$~Myr.
In the next 1~Myr, no significant changes occur in the comet's orbital elements.
During this quiet time (i.e., $1.3~{\rm Myr} < \tau < 2.3~{\rm Myr}$) the eccentricity
undergoes only small variations ({\it green} in Fig.~\ref{Fig3}).  In the same figure
we depict the perihelion distance ({\it blue} line) for this time interval showing
smooth changes from $q = 3.4$ to $5.5$~au; this is a safe zone void from any encounters.
At the end of this stage (i.e., $\tau = 2.3$~Myr), the comet approaches the planet
HD~10180~h again ($q = 3.4$), which transforms the comet's orbit into a chaotic one.

During the time interval ahead (i.e., $3.1~{\rm Myr} < \tau < 3.8~{\rm Myr}$), the
comet experiences only small changes regarding three dynamic properties, i.e.,
the semi-major axis $a$ ({\it blue} line in Fig.~\ref{Fig4}), the eccentricity $e$
({\it green} in Fig.~\ref{Fig3}), and the inclination $i$ ({\it green} line in
Fig.~\ref{Fig4}). 
After several close encounters with HD~10180~g (not shown here) visible through
large variations in all orbital parameters at approximately 3.9~Myr the comet enters
into a more quiet regime. Another feature resulting in a relatively stable orbit occurs
between 4.1~Myr and 4.8~Myr. In this case, its semi-major axis is relatively small
($a \simeq 3.0$) with increasing values towards the end of that period. 

Finally, we stop the integration after 5~Myr, a time when the inclination undergoes
big fluctuations between 25$^\circ$ and 55$^\circ$ for this comet. Detailed results
are given in Fig.~\ref{Fig3} and~\ref{Fig4}, with {\it red} marking the encounters
with planet HD~10180~h in Fig.~\ref{Fig3}. Although sometimes the comet approaches
the planet as close as ten planetary radii, no collisions are found to occur.  

%%%%%%%%%%%%%%%%%%%%%%%%%%%%%%%%%%%%%%%%%%%%%%%%%%%%%%%%%%%%%%%%%%%%%%%%%%%%%%

\section{Summary and Conclusions}
\label{sec:summary}

The purpose of this work is to explore the orbital dynamics of possible
exocomets in the HD~10180 star--planet system.  Exocomets appear to
constitute a significant component of stellar systems as informed by
previous theoretical and observational studies.  Besides the Solar System,
the best studied case so far is $\beta$~Pictoris \citep{kie14}, which
appears to host two families of comets of different histories and
degrees of fragmentation.  Additional findings have been reported for at
least 10 other stars \citep[e.g.,][]{zuc12,wel13}, with most of them being
relatively young and possibly in the stage of planet formation.
Continuing observational results are expected from the forthcoming
{\it Large Synoptic Survey Telescope}\footnote{See \url{https://www.lsst.org/.}}.

Previous theoretical work about the detectability of exocomets include
estimates by \cite{hai11}, who scaled realistic and extreme cases of the
Solar System to the distance of neighboring stars.  \cite{coo16} discussed
the detectability of close interstellar comets in consideration of previous
estimates by \cite{mor09}; this study is expected to be helpful for future
searches.  Generally speaking, the study of exocomets, either belonging to
the environments of stars or being interstellar in nature, carries
significant information about planet formation and the prospects
of habitable domains.  Current work mostly focuses on young A-type
stars \citep[e.g.,][]{zuc12,wel13}, but more comprehensive search
missions are anticipated in the foreseeable future.

In the present work, we consider the system of HD~10180.  The center star
is of spectral type G1~V, and is strikingly similar to the Sun regarding mass,
surface temperature, metallicity and age \citep[][and references therein]{lov11},
with some of the data even agreeing with solar values within their error bars.
Based on the work by \cite{lov11} and \cite{tuo12} we know that the HD~10180 system
is rich in planets as it may host up to nine planets (or even more, notwithstanding
future detections), which makes it the present-time record holder, thus surpassing
the Solar System.  However, the planetary system of HD~10180 is much more compact
than the Solar System.  The outermost planet HD~10180~h has a distance from its
star comparable to the big gap for Solar System asteroids between the 2:1 and 3:2
mean motion resonance with Jupiter.

\begin{figure*}
\centering
\begin{tabular}{c}
\epsfig{file=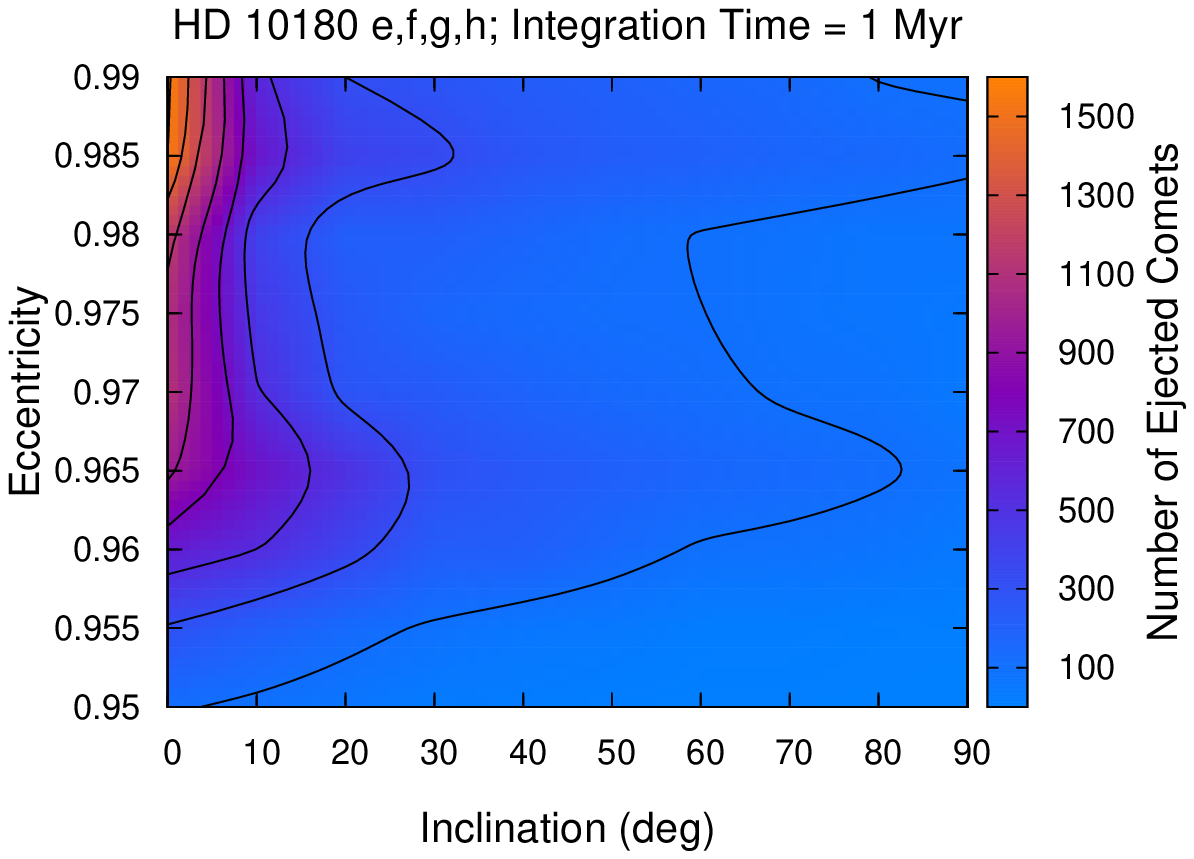,width=0.45\textwidth}
\epsfig{file=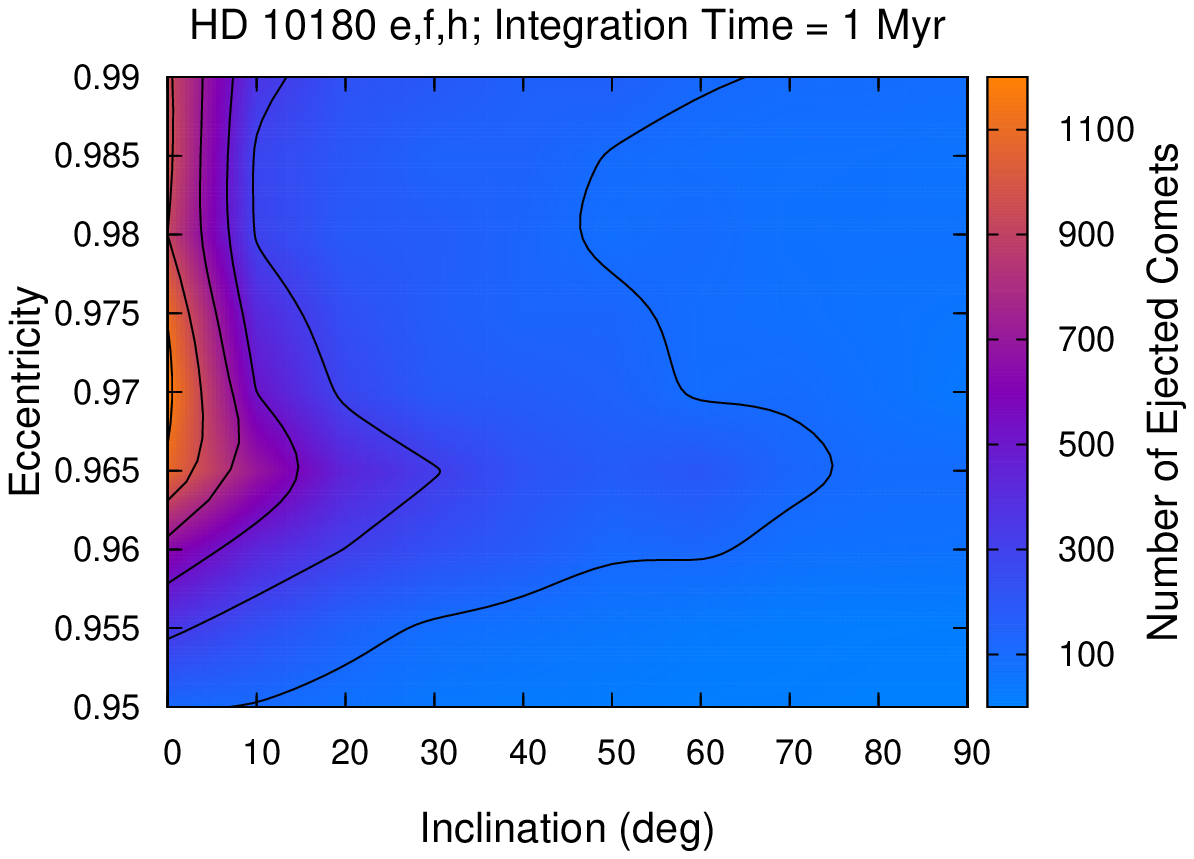,width=0.45\textwidth} \\
\epsfig{file=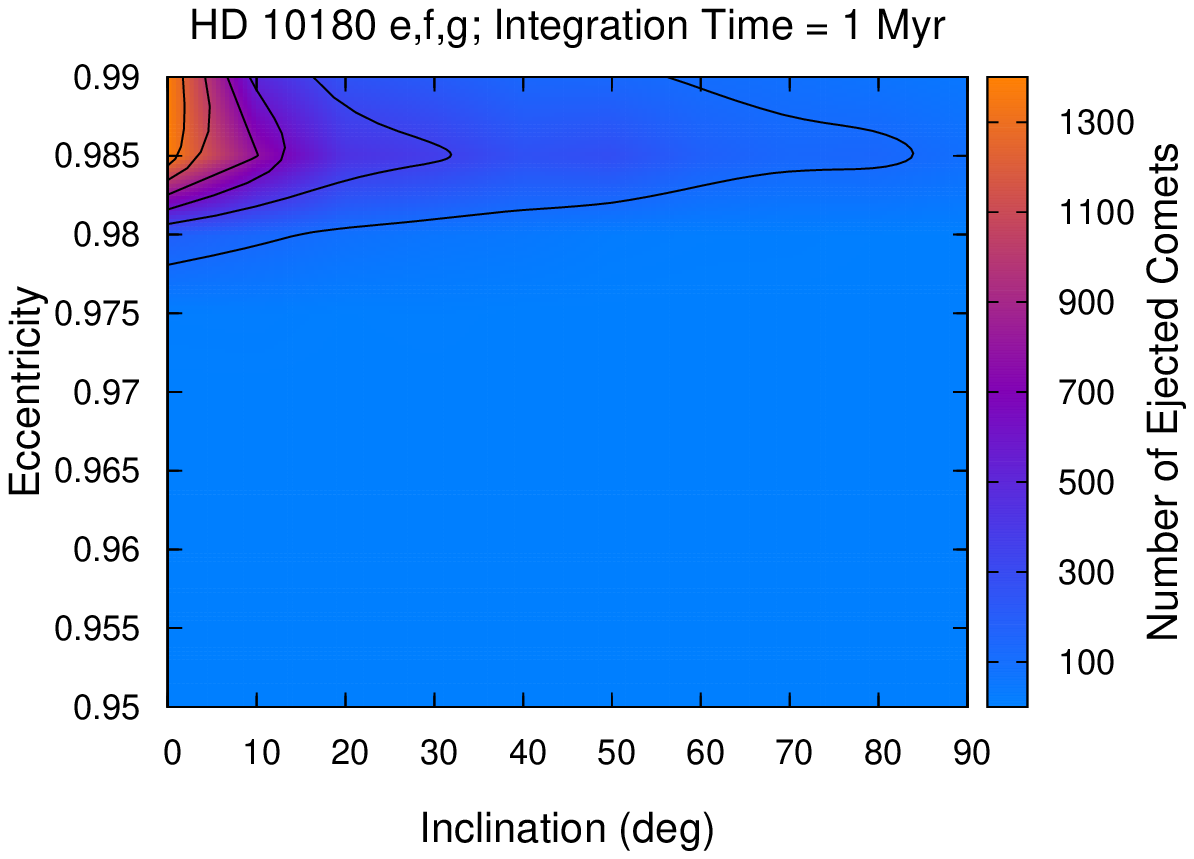,width=0.45\textwidth}
\epsfig{file=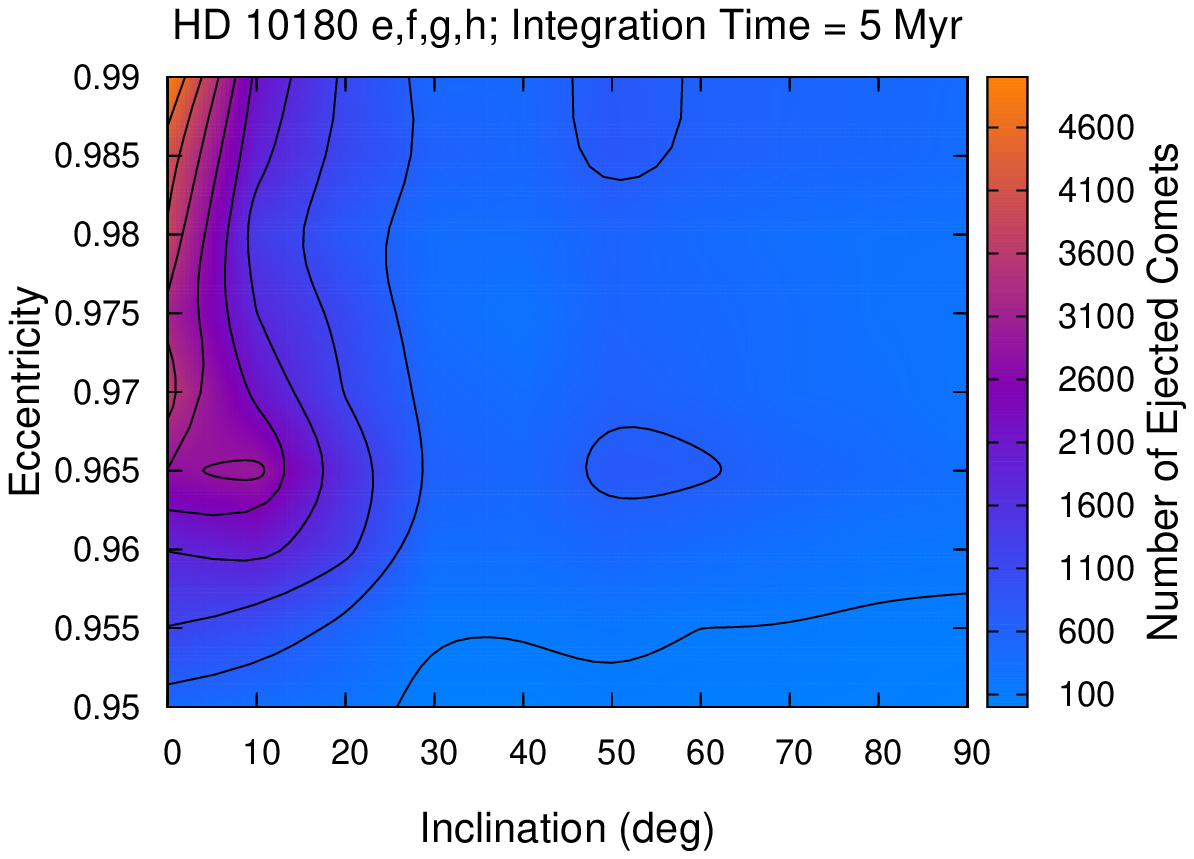,width=0.45\textwidth}
\end{tabular}
\caption{The color scheme indicates the number of escaped comets from the system
for different initial conditions $(e_0, i_0)$ and an integration time
of 1~Myr for the different setups used:
{\it Upper left:} Same as if Fig.~\ref{Fig5}, but here only shown for comets with
initial inclination of up to $90^{\circ}$.  The orange color in the left upper corner
indicates that most of the comets with high eccentricity and low initial inclination
are ejected from the system.
{\it Upper right:} The exclusion of the second massive planet HD~10180~g does not
really change the picture; see comparison to the {\it upper left} panel.
The total number of comets ejected from the system is slightly smaller.  Furthermore,
the orange regime is shifted to comets with lower initial eccentricity.
{\it Lower left:} The exclusion of planet HD~10180~h has a big influence on the
outcome.  Comets with initial eccentricity lower than 0.98 are no longer affected
by interactions with the planets and thus stay in the system.  Only the ones with
initial eccentricities of $> 0.98$ reach the orbit of planet g and thus can be ejected
from the system.
{\it Lower right:} The calculation of the whole system (with 4 planets) for an
integration time of 5~Myr shows almost the same pattern as the two panels in the
{\it upper} row.  The total number of ejected comets is higher, however, as expected.
}
\label{Fig8}
\end{figure*}

\cite{ste90} proposed that based on the detection rate of interstellar comets
one could draw conclusions about the rate of planetary formation in the galaxy.
In our study, we did not investigate the influence of planet formation or, e.g.,
the impact of giant planet migration on the comet ejection rates.
Nevertheless, we found that due to gravitational interaction with the planets
a substantial number of comets were ejected from the system, which allows us
to conclude that not only planet formation mechanisms, but also the influence
of passing stars with ensuing interactions of the comets with the planets and
subsequent ejection from the system can account for a large number of interstellar
comets (e.g., see Fig.~\ref{Fig8} {\it lower right panel}).

Our simulations also showed that planet HD~10180~h dominates the
cometary scattering process depending on the comets' initial conditions. It is also governing the orbital dynamics of exocomets
considered in our study; a feature also shared by Jupiter in the Solar System.
Our simulations took into account more than 30,000 comets as part of the
initial orbital integration scheme, which included comets of different
initial eccentricity and inclination (also encompassing comets in retrograde
orbits).  Our study shows different kinds of outcomes such as encounters with
the planets, captures, escapes, and secular comet--planet resonances.  The
integration process itself included the four outer planets of the system,
i.e., planet e, f, g, and h (see Table~\ref{Table1}).  Generally, comets
with relatively large eccentricities were able to enter the inner region
of the system, thus facing early close planetary encounters.  Owing to our
theoretical approach, the number of ejected comets considered in our simulation
readily increased over time.

\begin{figure*}
\centering
\begin{tabular}{c}
\epsfig{file=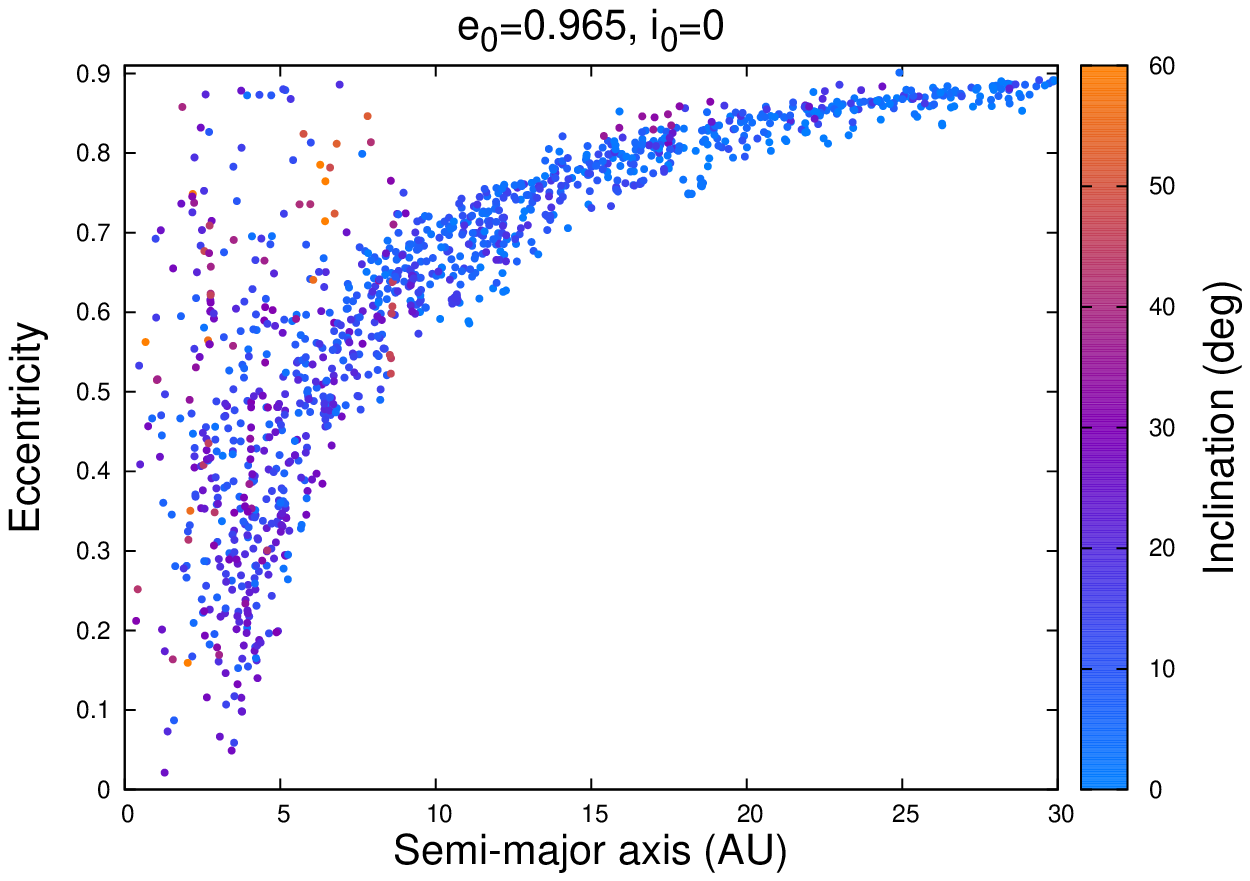,width=0.5\textwidth}
\epsfig{file=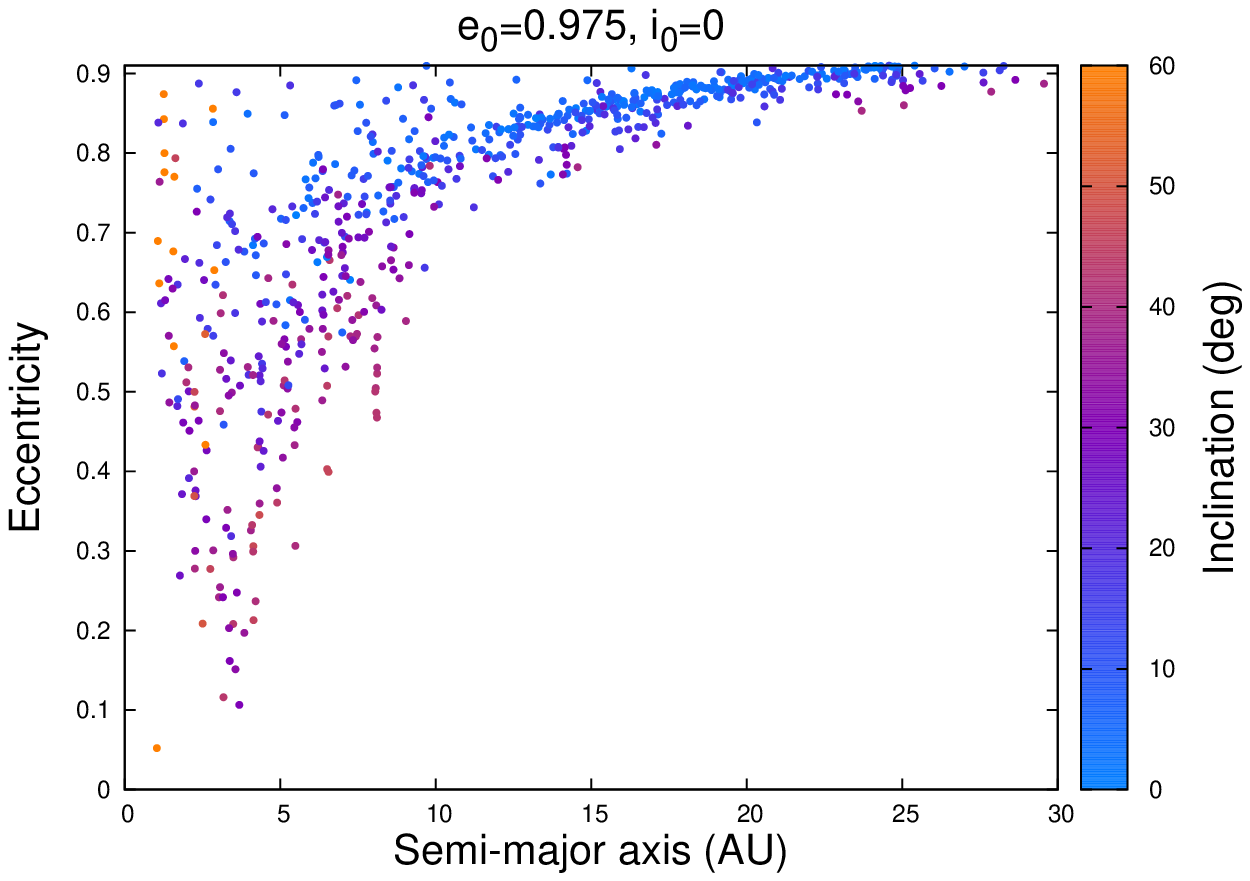,width=0.5\textwidth}\\
\epsfig{file=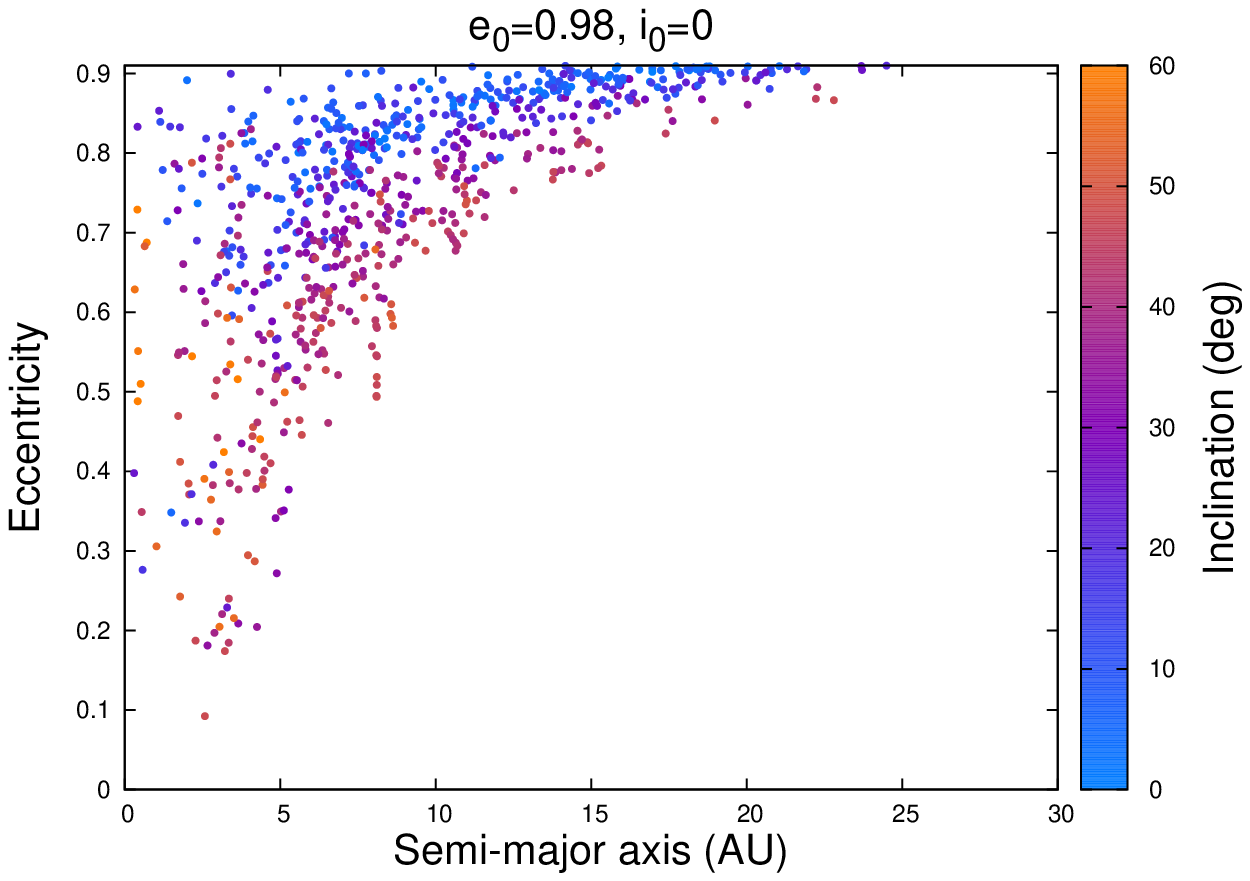,width=0.5\textwidth}
\epsfig{file=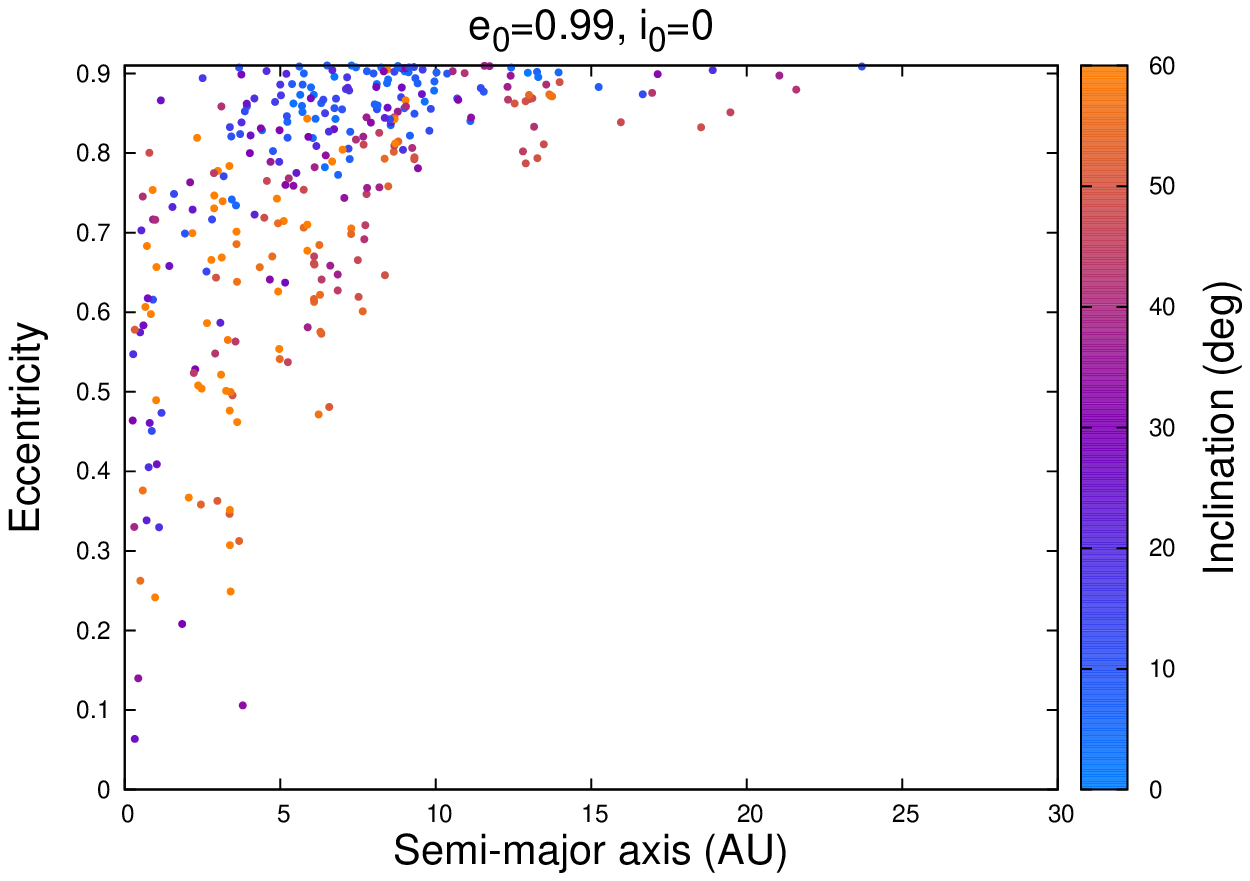,width=0.5\textwidth}
\end{tabular}
\caption{Orbital properties of captured comets with four different initial
eccentricities, which are $e_0=0.965$, 0.975, 0.98, and 0.99, and an initial inclination
of $i_0\,=\,0^{\circ}$. The color scheme indicates the inclination at the time of capture.
For these computations all four planets have been included and the integration time has
been set to 1 Myr.  Our results indicate that some comets are captured in orbits of low
eccentricity and a small semi-major axis.  Those comets could potentially form a
\textit{Jupiter family} analogue in the HD~10180 system.  However, no clear distinction
between short- and long-period comets could be identified.  Nevertheless, most comets
are captured in highly eccentric orbits and will probably be ejected from the system
later on.  The higher the initial eccentricity $e_0$, the less probable it is for a comet
to be captured in an orbit with moderate values for its semi-major axis and eccentricity.
Interestingly, some comets, especially those with large initial eccentricity
($e_0 \, > \, 0.97 \,$), are scattered to inclinations of up to $60^{\circ}$.
}
\label{Fig11}
\end{figure*}

\begin{figure*}
\centering
\begin{tabular}{c}
\epsfig{file=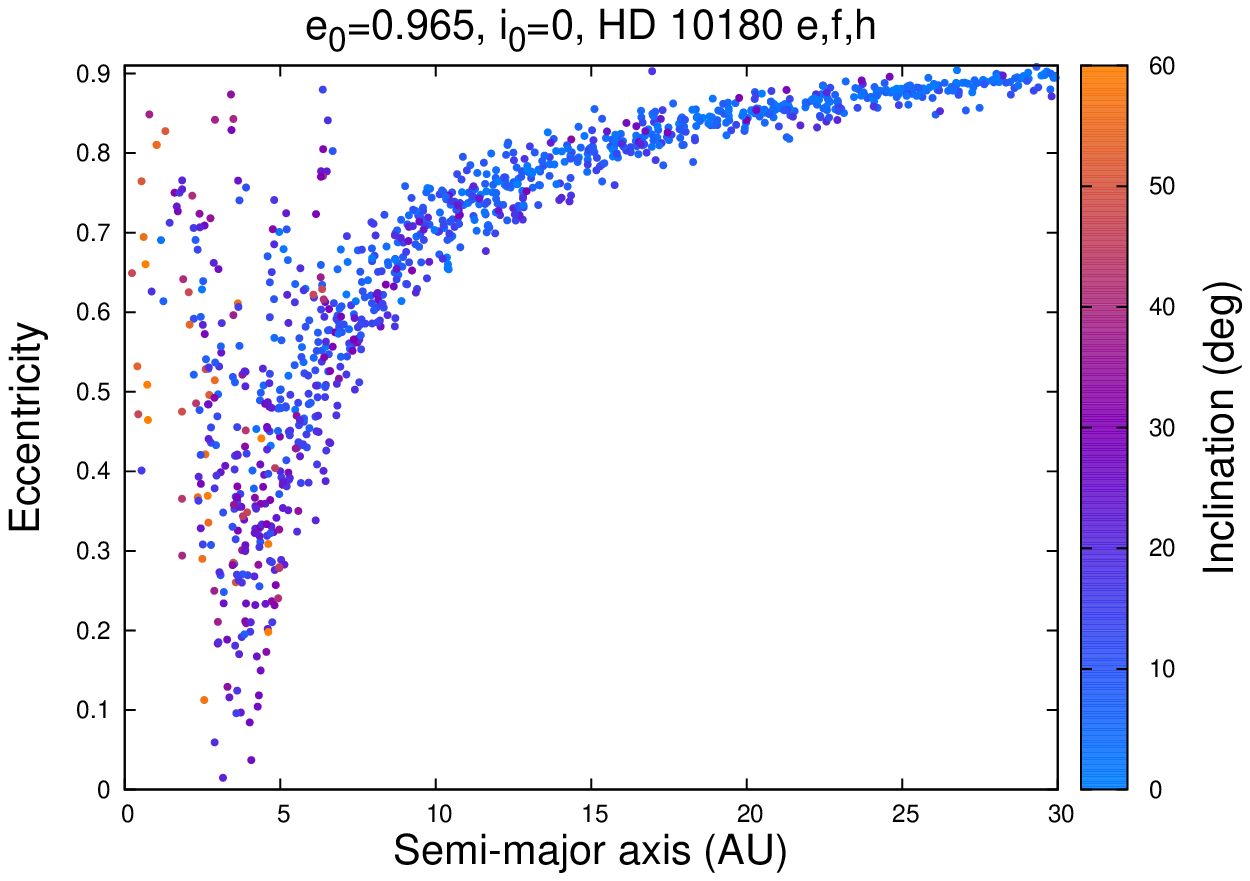,width=0.5\textwidth}
\epsfig{file=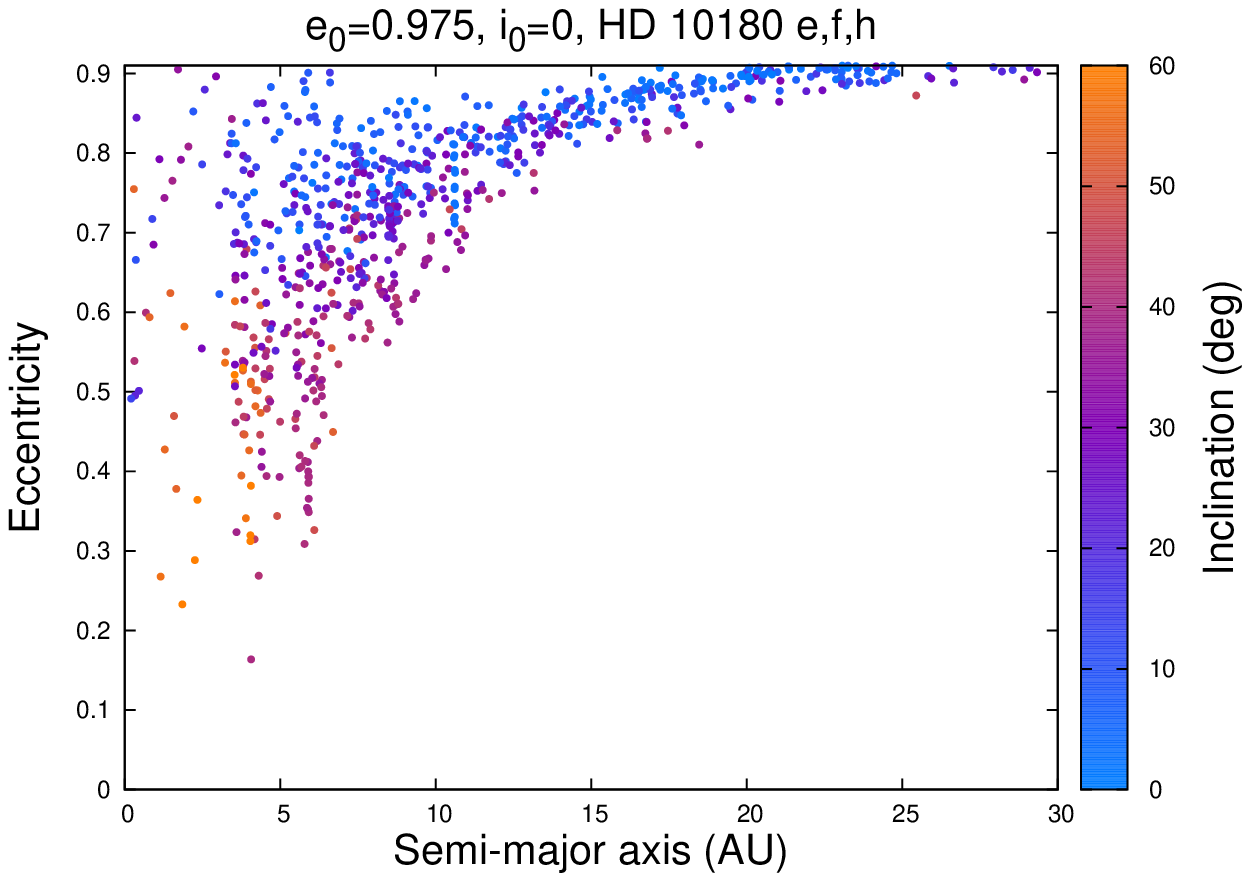,width=0.5\textwidth}\\
\epsfig{file=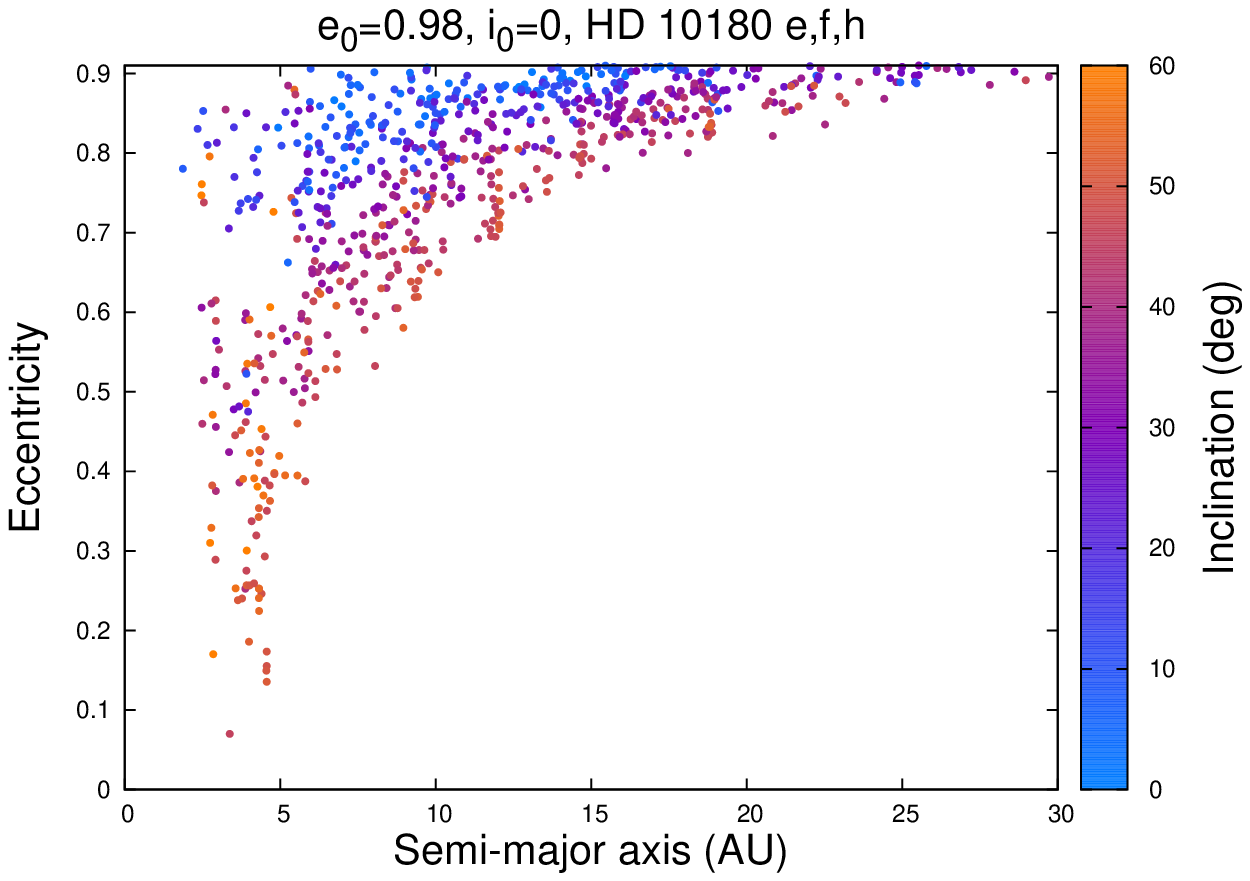,width=0.5\textwidth}
\epsfig{file=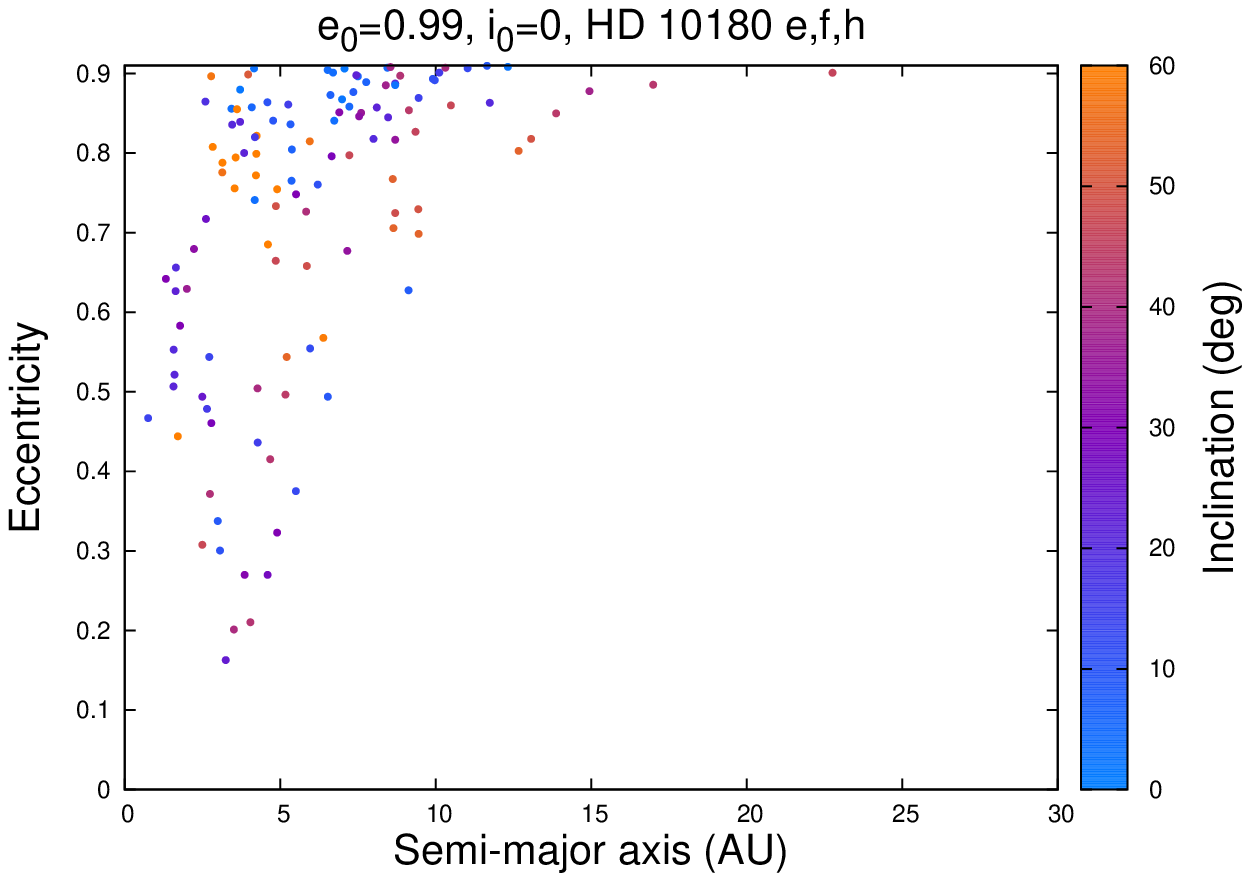,width=0.5\textwidth}
\end{tabular}
\caption{Same as Fig.~\ref{Fig11}, but with planet HD~10180~g excluded
from the system. Comparing this figure with Fig.~\ref{Fig11} with all planets
included reveals similar outcomes, indicating that the influence of planet g
on the orbital properties of captured comets is relatively small.  Again, comets
captured on close in orbits may have high inclinations, especially for comets with $e_0 < 0.97$.
}
\label{Fig12}
\end{figure*}

\begin{figure*}
\centering
\begin{tabular}{c}
\epsfig{file=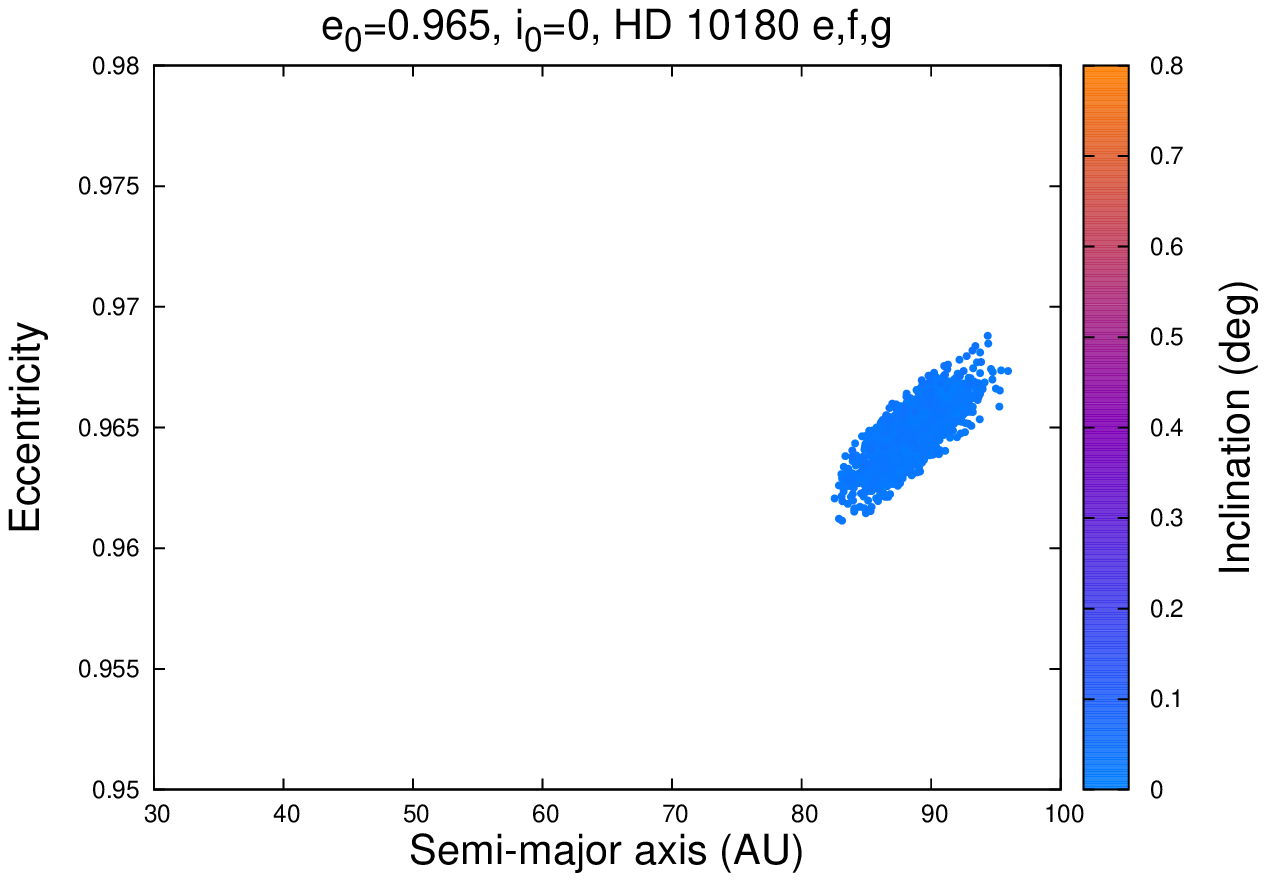,width=0.5\textwidth}
\epsfig{file=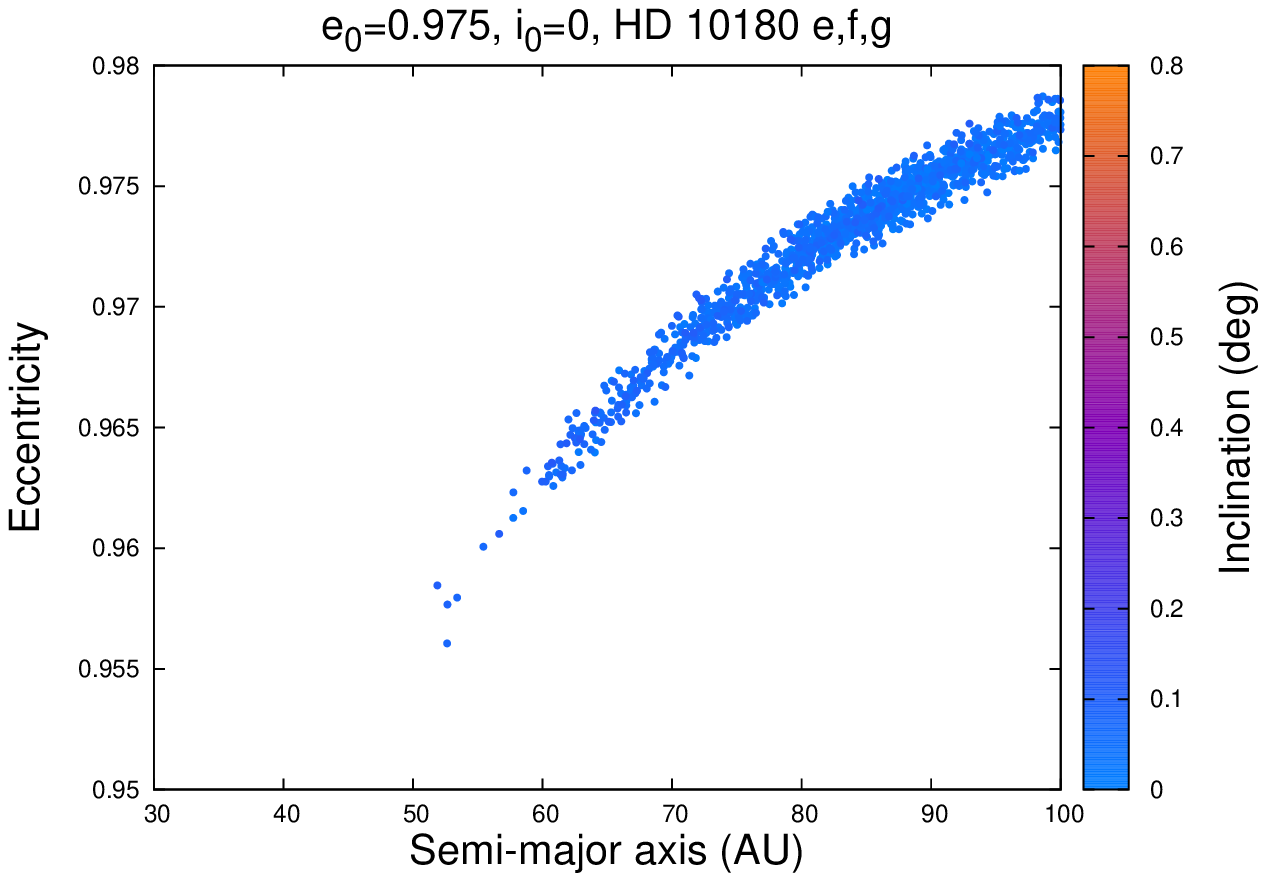,width=0.5\textwidth}\\
\epsfig{file=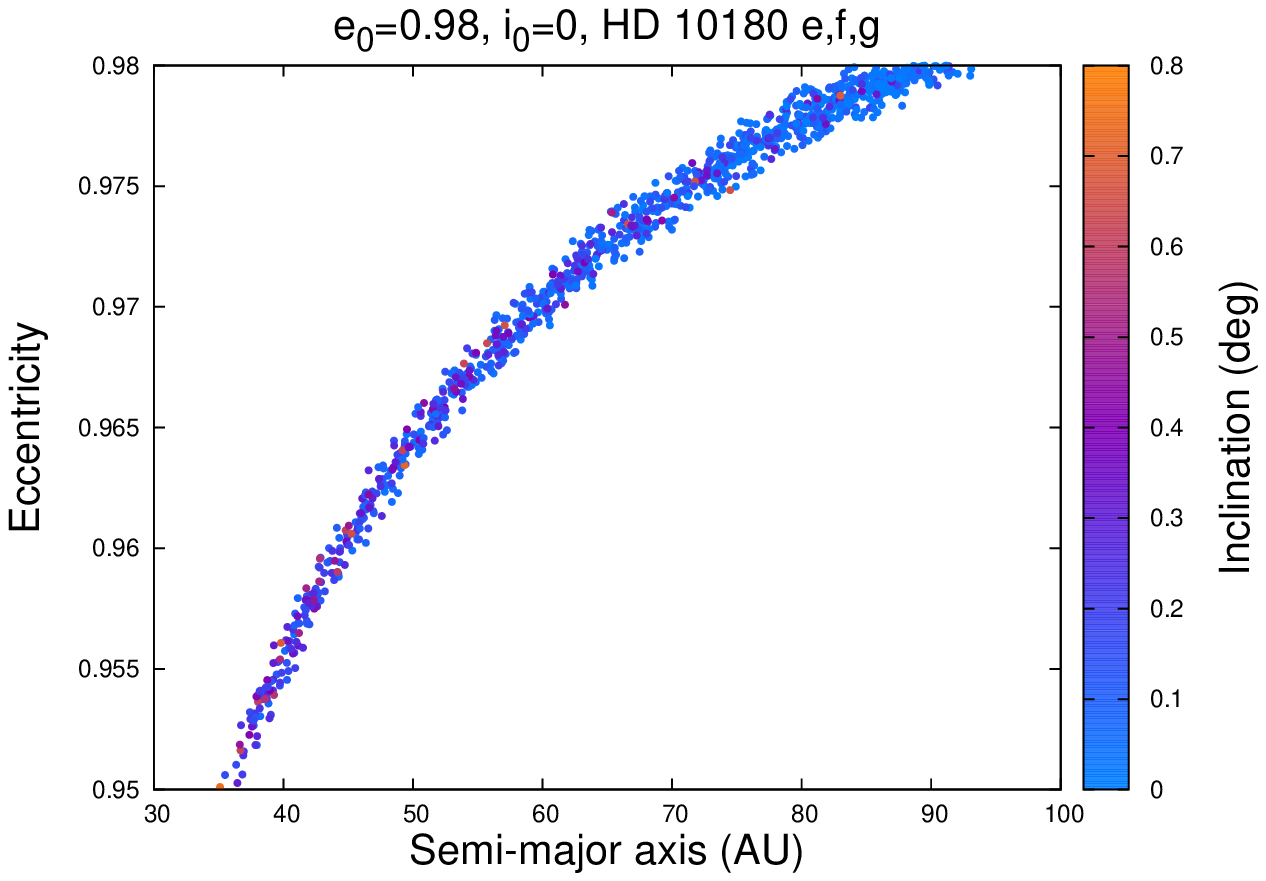,width=0.5\textwidth}
\epsfig{file=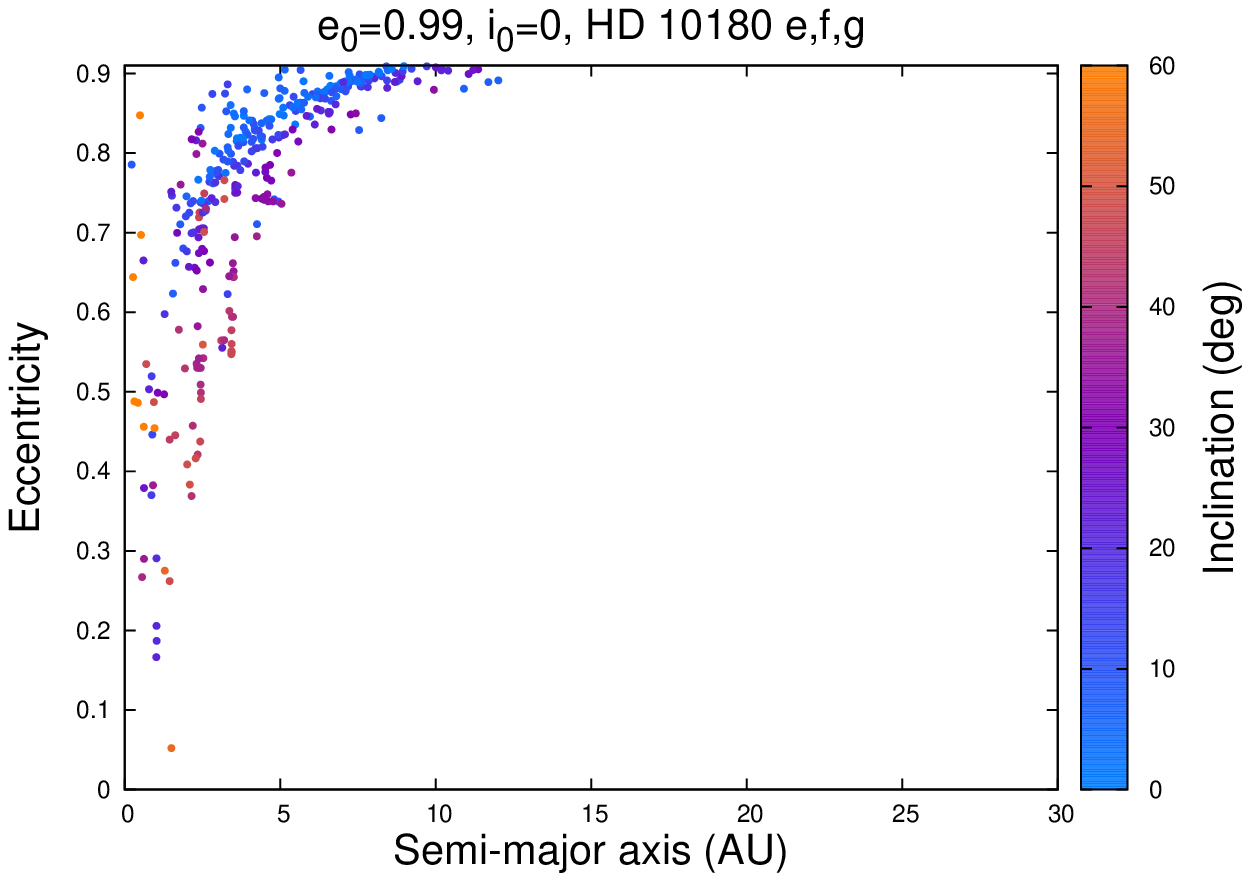,width=0.5\textwidth}
\end{tabular}
\caption{Same as Fig.~\ref{Fig11}, but with planet HD~10180~h (the most
massive planet) excluded from the system.  The outcome is now significantly different
from the simulations with all four planets considered.  Comets with initially small
eccentricity do not reach the orbit of the now most massive planet of the system,
HD~10180~g.  Thus, they cannot interact and their orbits remain unchanged.
For example, the upper left panel, pertaining to $e_0 = 0.965$, indicates that
comets still populate orbits with high eccentricities and semi-major axes of
$a > 80\,$au akin to their initial values. (Note the range of the $y$-axis!)
The picture changes slightly with increasing initial eccentricities, which allows
comets to penetrate into the inner planetary system and to start interacting with
planet HD~10180~g.  For comets with initial eccentricity of $e_0 = 0.98$, it is possible
for them to be captured on orbits with semi-major axes $a \, < \, 40$~au, but the
eccentricity for these orbits still remains high, which makes a permanent capture
unlikely.  Based on the result of this figure with planet HD~10180~h omitted, as
well as the comparisons to Fig.~\ref{Fig11} and Fig.~\ref{Fig12}, we can conclude
again that planet HD~10180~h definitely plays the main role in capturing comets.
}
\label{Fig13}
\end{figure*}

\begin{figure*}
\centering
\begin{tabular}{c}
\epsfig{file=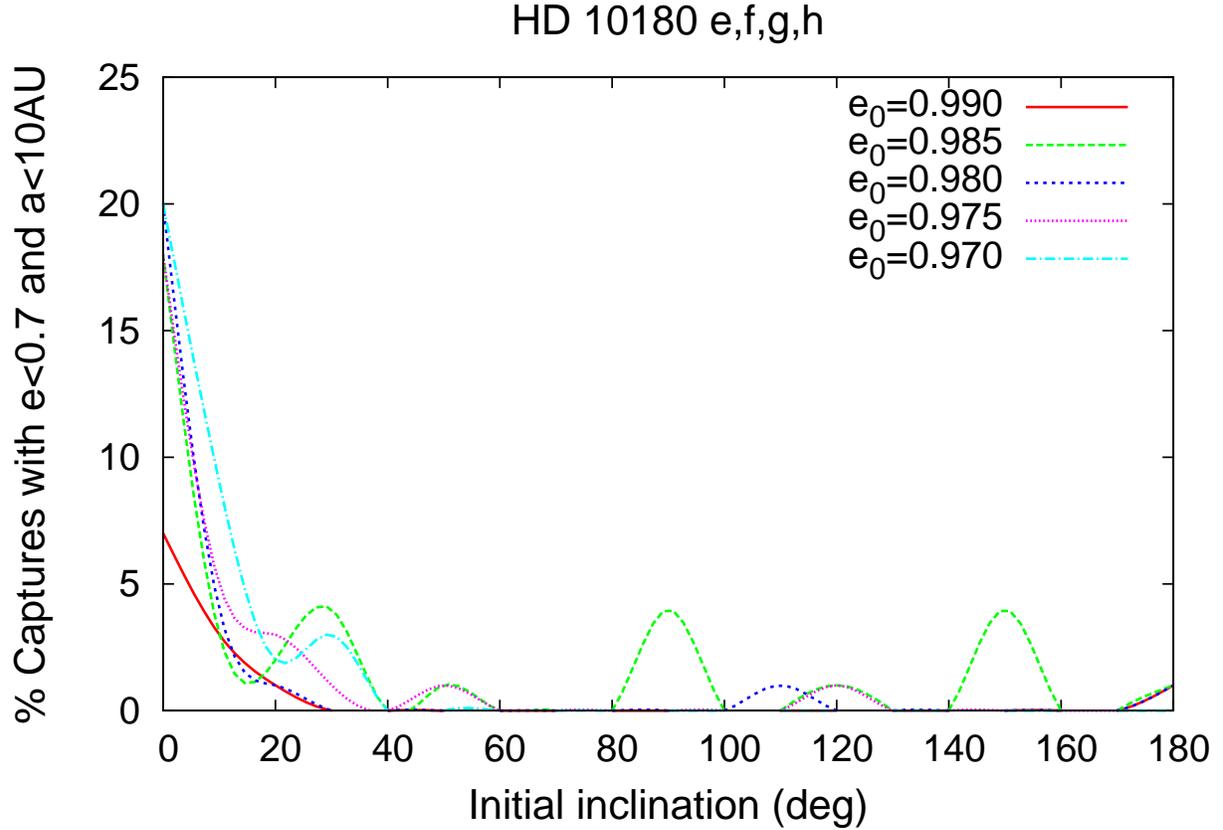,width=1.0\textwidth}
\end{tabular}
\caption{Probability for a comet with specific initial conditions
($e_0$,$i_0$) to be captured in an orbit with $e \, < \, 0.7 $ and $a \, < \, 10 \, $~au.
The highest probability of capture occurs for an initial eccentricity between
${e_0}=$~0.97 and 0.98 combined with an initial inclination close to $0^{\circ}$.
The next highest peaks arise for ${i_0}=30^{\circ}$, ${i_0}=90^{\circ}$, and ${i_0}=150^{\circ}$.  These occurrences mark
initial inclinations largely coinciding with the system's plane of orbit and
perpendicular to that plane. The probability for a comet with initial eccentricity
lower than ${e_0}=$~0.97 is close to zero for all inclinations.  Note that comets of
${e_0}=0.975$ have the highest probability to be captured.
}
\label{Fig14}
\end{figure*}

\begin{figure*}
\centering
\begin{tabular}{c}
\epsfig{file=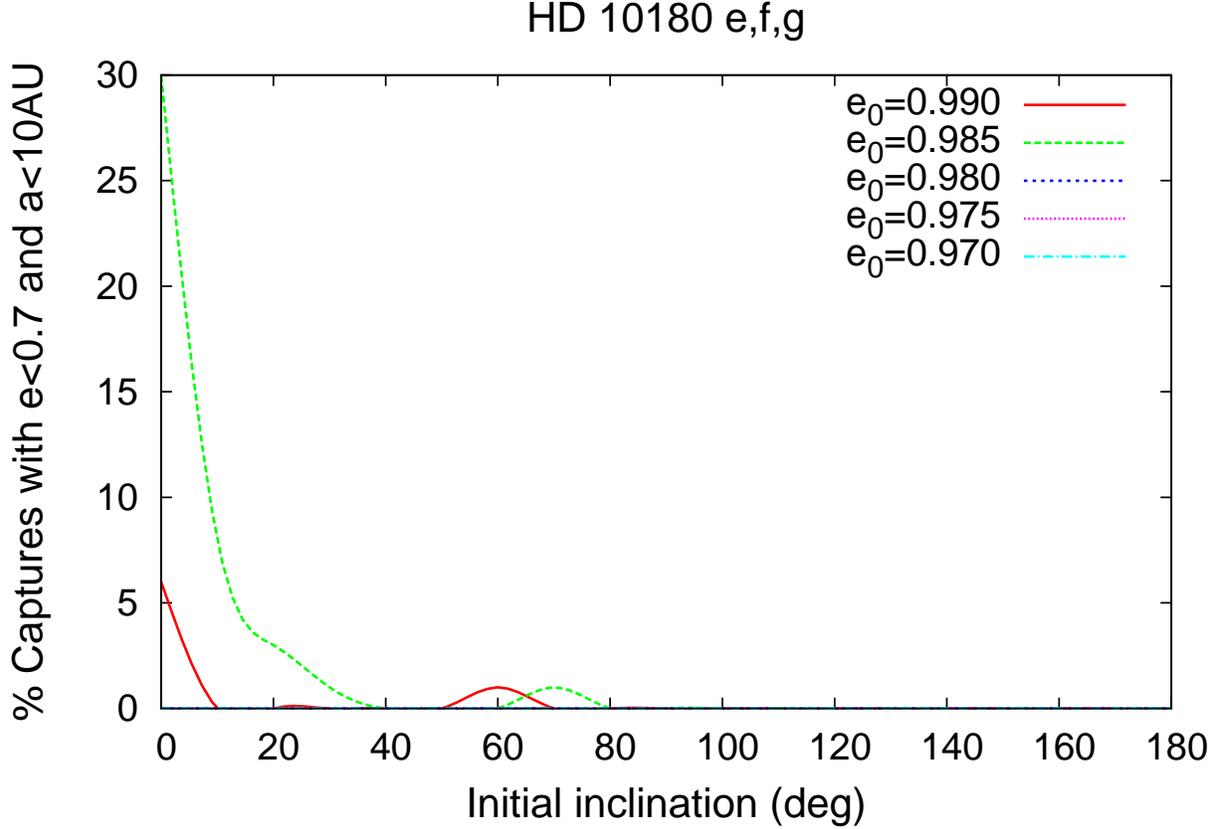,width=1.0\textwidth}
\end{tabular}
\caption{Same as Fig.~\ref{Fig14}, but without planet HD~10180~h. Interestingly,
the probability for a capture into orbits with moderate values for $a$ and $e$ is 30$\%$
for comets with initial eccentricities of 0.985.  This percentage is even higher than
for models with planet h included (compare Fig.~\ref{Fig14}).  Thus, we can conclude
that the combination of the scattering given by planet g and h plays a prevalent role
for comets with this particular initial eccentricity.  Noting that the probability for
a capture for comets with $e_0$=0.99 remains about the same, leads to the conclusion
that for comets in highly eccentric orbits planet g plays the main role in the scattering
process.  On the other hand, comets with lower initial eccentricities, i.e., $e < 0.985$,
have a perihelion distance beyond the orbit of planet g and thus cannot be majorly affected
by this planet, which is why the probability for them to be captured is zero.}
\label{Fig15}
\end{figure*}

It was also found that for comets with large eccentricities many more escapes
from the system occurred.  However, the border between completely stable and
escaping cometary orbits is not a sharp one; in fact, at the interface separating
the two regions, there is a small band characterized by statistic fluctuations.
Moreover, there are no ejections for retrograde orbits up to eccentricities of
$e_0=0.94$.  Nonetheless, even without crossings of cometary orbits with any of
the planets, the impact of secular resonances can lead to changes in cometary
orbits so that close planetary encounters are possible.  We also showed how comets
of initially planar orbits, in the context of our simulations, are spread evenly
to also include retrograde orbits and could also be in long-term stable orbits
around the host star (see Fig.~\ref{Fig11}).

Even though the idea of finding families of comets in the HD~10180 system was
a key motivation for this work, we were unable to determine any dynamical families,
although a very large number of captured comets has been identified through our
numerical integrations.  
We think that the reason why no families of comets in the HD~10180 system
were found is because of the system's architecture, which aside from the
many affinities to the Solar System still exhibits decisive differences.
For example, in our system beyond the most massive planet Jupiter other massive
planets exist able to influence the orbits of incoming objects,
whereas in the HD~10180 system, the most massive planet h constitutes the system's
outermost planet.

A particular focus of our work concerned the role of the
comet's initial eccentricity and inclination regarding capture, evolution
of orbital elements, or the comet's escape.  In the future, we plan to extend
our studies of comets to other exoplanetary systems including systems with
compositions akin to the Solar System as well as systems with notably
different structures.  Moreover,
we consider our work as part of the bigger picture associated with the study
of comets, pertaining to the Solar System as well as to exosolar systems.
Studies about exocomets yield a large range of implications, involving the
origin, structure and evolution of systems as well as the proliferation of
water and prebiotic compounds to terrestrial planets, which will increase
their chances of  being habitable.  This latter aspect has been showcased
by results from {\it Rosetta} \citep[e.g.,][]{cap15,ric15,bos16}, which indicate
the pivotal role of comets regarding prebiotic chemistry and potential
exobiology.

To consider the importance of exocomets for possible life in exosolar systems
another possible extension of our future work will include the detailed study
of encounters and collisions with planet HD~10180~g located at the outer edge
of the system's habitable zone, comparable to the position of Mars in the Solar System.  This
exosolar planet is Neptune-like; therefore, not being able to host life.
However, an exomoon or a Trojan-type object associated with HD~10180~g could
still be habitable.

%%%%%%%%%%%%%%%%%%%%%%%%%%%%%%%%%%%%%%%%%%%%%%%%%%%%%%%%%%%%%%%%%%%%%%%%%%%%%%

\acknowledgments
This research is supported by the Austrian Science Fund (FWF) through
grant S11603-N16 (B. L. and R. D.).  Moreover, M. C. acknowledges support
by the University of Texas at Arlington. We also thank an unknown referee
for valuable advice allowing us to improve this paper.

%%%%%%%%%%%%%%%%%%%%%%%%%%%%%%%%%%%%%%%%%%%%%%%%%%%%%%%%%%%%%%%%%%%%%%%%%%%%%%

\clearpage

%\bibliographystyle{natbib}

%%%%%%%%%%%%%%%%%%%%%%%%%%%%%%%%%%%%%%%%%%%%%%%%%%%%%%%%%%%%%%%%%%%%%%%%

\clearpage

%%% *** Fig.1
%%%%%%%%%%%%%%%%%%%%%%%%%%%%%%%%%%%%%%%%%%%%%%%%%%%%%%%%%%%%%%%%%

%+++++++++++++++++++++++++++++++++++++++++++++++++++++++++++++++++++++++

\clearpage

%%% *** Fig.2
%%%%%%%%%%%%%%%%%%%%%%%%%%%%%%%%%%%%%%%%%%%%%%%%%%%%%%%%%%%%%%%%%

%+++++++++++++++++++++++++++++++++++++++++++++++++++++++++++++++++++++++

\clearpage

%%% *** Fig.3
%%%%%%%%%%%%%%%%%%%%%%%%%%%%%%%%%%%%%%%%%%%%%%%%%%%%%%%%%%%%%%%%%

%+++++++++++++++++++++++++++++++++++++++++++++++++++++++++++++++++++++++

\clearpage

%%% *** Fig.4
%%%%%%%%%%%%%%%%%%%%%%%%%%%%%%%%%%%%%%%%%%%%%%%%%%%%%%%%%%%%%%%%%

%+++++++++++++++++++++++++++++++++++++++++++++++++++++++++++++++++++++++

\clearpage

%%% *** Fig.5
%%%%%%%%%%%%%%%%%%%%%%%%%%%%%%%%%%%%%%%%%%%%%%%%%%%%%%%%%%%%%%%%%

%+++++++++++++++++++++++++++++++++++++++++++++++++++++++++++++++++++++++

\clearpage

%%% *** Fig.6
%%%%%%%%%%%%%%%%%%%%%%%%%%%%%%%%%%%%%%%%%%%%%%%%%%%%%%%%%%%%%%%%%

%+++++++++++++++++++++++++++++++++++++++++++++++++++++++++++++++++++++++

\clearpage

%%% *** Fig.7
%%%%%%%%%%%%%%%%%%%%%%%%%%%%%%%%%%%%%%%%%%%%%%%%%%%%%%%%%%%%%%%%%

%+++++++++++++++++++++++++++++++++++++++++++++++++++++++++++++++++++++++

\clearpage

%%% *** Fig.8
%%%%%%%%%%%%%%%%%%%%%%%%%%%%%%%%%%%%%%%%%%%%%%%%%%%%%%%%%%%%%%%%%

%+++++++++++++++++++++++++++++++++++++++++++++++++++++++++++++++++++++++
%\clearpage
%
%%%% *** Fig.9
%%%%%%%%%%%%%%%%%%%%%%%%%%%%%%%%%%%%%%%%%%%%%%%%%%%%%%%%%%%%%%%%%%
%\begin{figure*}
%\centering
%\begin{tabular}{c}
%\epsfig{file=Fig9.eps,width=0.45\textwidth}
%\end{tabular}
%\caption{The color scheme indicates the number of escaped comets from the system
%for different initial conditions $(e_0,i_0)$ and an integration time
%of 1~Myr with the planet HD~10180~h excluded from the system (the most
%massive planet).  The outcome is now significantly different from the
%simulations with all four planets considered; see Fig.~\ref{Fig5}.
%Here only comets with initial eccentricity $e_0 \, > \, 0.98$ and
%low initial inclinations are scattered to be ejected from the system.
%We are thus able to conclude that planet h is the most important
%perturber in the system.
%}
%\label{Fig9}
%\end{figure*}
%
%%+++++++++++++++++++++++++++++++++++++++++++++++++++++++++++++++++++++++
%\clearpage
%
%%%% *** Fig.10
%%%%%%%%%%%%%%%%%%%%%%%%%%%%%%%%%%%%%%%%%%%%%%%%%%%%%%%%%%%%%%%%%%
%\begin{figure*}
%\centering
%\begin{tabular}{c}
%\epsfig{file=Fig10.eps,width=1.0\textwidth}
%\end{tabular}
%\caption{Same as Fig.~\ref{Fig5}, but for an integration time of 5~Myr.
%All four system planets have been included.  The figure shows that
%the outcomes of the scattering process do not differ strongly from
%the ones obtained after 1 Myr.  The only notable difference is that
%a larger number of comets has been ejected from the system.
%}
%\label{Fig10}
%\end{figure*}
%
%+++++++++++++++++++++++++++++++++++++++++++++++++++++++++++++++++++++++
\clearpage

%%% *** Fig.11
%%%%%%%%%%%%%%%%%%%%%%%%%%%%%%%%%%%%%%%%%%%%%%%%%%%%%%%%%%%%%%%%%

%+++++++++++++++++++++++++++++++++++++++++++++++++++++++++++++++++++++++
\clearpage

%%% *** Fig.12
%%%%%%%%%%%%%%%%%%%%%%%%%%%%%%%%%%%%%%%%%%%%%%%%%%%%%%%%%%%%%%%%%

%+++++++++++++++++++++++++++++++++++++++++++++++++++++++++++++++++++++++

\clearpage

%%% *** Fig.13
%%%%%%%%%%%%%%%%%%%%%%%%%%%%%%%%%%%%%%%%%%%%%%%%%%%%%%%%%%%%%%%%%

%+++++++++++++++++++++++++++++++++++++++++++++++++++++++++++++++++++++++

\clearpage

%%% *** Fig.14
%%%%%%%%%%%%%%%%%%%%%%%%%%%%%%%%%%%%%%%%%%%%%%%%%%%%%%%%%%%%%%%%%

%+++++++++++++++++++++++++++++++++++++++++++++++++++++++++++++++++++++++
\clearpage

%%% *** Fig.15
%%%%%%%%%%%%%%%%%%%%%%%%%%%%%%%%%%%%%%%%%%%%%%%%%%%%%%%%%%%%%%%%%

%+++++++++++++++++++++++++++++++++++++++++++++++++++++++++++++++++++++++

\clearpage

% \end{document}

%%% *** Table 1
%%%%%%%%%%%%%%%%%%%%%%%%%%%%%%%%%%%%%%%%%%%%%%%%%%%%%%%%%%%%%%%%%%%%%%%%%
\begin{deluxetable}{lllcccrl}
\tablecaption{The HD~10180 Planetary System, Part I}
\tablewidth{0pt}
\tablehead{
Name & $a$ & $e$ & $i$ & $\omega$  & $\Omega$ & $M$ & $m$ \\
\noalign{\smallskip}
\hline
\noalign{\smallskip}
 ... & (au) &  ...  &  ($^\circ$) & ($^\circ$) & ($^\circ$) & ($^\circ$) & ($M_\odot$)     
}
\startdata
HD~10180~e &  0.270   & 0.0260  &  0.70 & 1.0 & 1.0 &  50.0 & 0.0000790 \\
HD~10180~f &  0.49220 & 0.1350  &  0.70 & 1.0 & 1.0 &  90.0 & 0.0000750 \\
HD~10180~g &  1.4220  & 0.00010 &  0.80 & 1.0 & 1.0 & 181.0 & 0.0000670 \\
HD~10180~h &  3.40    & 0.080   &  0.60 & 1.0 & 1.0 &   1.0 & 0.0002    \\
\enddata
\tablecomments{
The four outer planets of HD~10180, which have been included in our study.
Moreover, there are three additional planets orbiting inside of HD~10180~e,
as well as two unconfirmed planets; see \cite{tuo12} and
{\tt http://www.exoplanet.eu} for details and updates.}
\label{Table1}
\end{deluxetable}

%+++++++++++++++++++++++++++++++++++++++++++++++++++++++++++++++++++++++

\clearpage

%%% *** Tab.2
%%%%%%%%%%%%%%%%%%%%%%%%%%%%%%%%%%%%%%%%%%%%%%%%%%%%%%%%%%%%%%%%%%%%%%%%%
\begin{deluxetable}{lllccc}
\tablecaption{The HD~10180 Planetary System, Part II}
\tablewidth{0pt}
\tablehead{
Name & Mass & $R_\mathrm{Hill}$ & $R_\mathrm{Hill}$  & $r_{\rho_1}$   & $r_{\rho_2}$ \\
\noalign{\smallskip}
\hline
\noalign{\smallskip}
 ... & ($m_\mathrm{Jupiter}$) &  (au)  &  ($10^6$ km) & ($10^3$ km) & ($10^3$ km)   
}
\startdata
HD~10180~e &  0.0827 & 0.0080 &  1.18 & 33 & 26   \\
HD~10180~f &  0.0786 & 0.0144 &  2.12 & 32 & 26   \\
HD~10180~g &  0.0702 & 0.0400 &  5.91 & 31 & 25   \\
HD~10180~h &  0.2095 & 0.1379 & 20.40 & 45 & 36   \\
\enddata
\tablecomments{
Masses (in $m_\mathrm{Jupiter}$), Hill radii and planetary radii (for two different densities)
for the four planets of HD~10180 considered in our integrations with assumed densities of
$\rho_1$ = 1~g~cm$^{-3}$ and $\rho_2$ = 2~g~cm$^{-3}$.}
\label{Table2}
\end{deluxetable}

%+++++++++++++++++++++++++++++++++++++++++++++++++++++++++++++++++++++++

\clearpage

%%% *** Tab.3
%%%%%%%%%%%%%%%%%%%%%%%%%%%%%%%%%%%%%%%%%%%%%%%%%%%%%%%%%%%%%%%%%%%%%%%%%
\begin{deluxetable}{rrrrrrrrr}
\tablecaption{Results for Comets in Prograde Orbits, Part~1}
\tablewidth{0pt}
\tablehead{
Incl.  & \multicolumn{8}{c}{Eccentricity}  \\
\hline
\noalign{\smallskip}
...         & 0.905 & 0.910 & 0.915 & 0.920 & 0.925 & 0.930 & 0.935 & 0.940
}
\startdata
  0$^\circ$ &     0 &     0 &     0 &     6 &    14 &    52 &   104 &  300 \\
 10$^\circ$ &     0 &     0 &     0 &     4 &     7 &    70 &    91 &  230 \\
 20$^\circ$ &     0 &     0 &     0 &     5 &     9 &    28 &    79 &  162 \\
 30$^\circ$ &     0 &     0 &     0 &     0 &     7 &    29 &    71 &  128 \\
 40$^\circ$ &     0 &     0 &     0 &     1 &     1 &    17 &    32 &   51 \\
 50$^\circ$ &     0 &     0 &     0 &     0 &     0 &     0 &     0 &    0 \\
 60$^\circ$ &     0 &     0 &     0 &     0 &     0 &     0 &     0 &    0 \\
 70$^\circ$ &     0 &     0 &     0 &     0 &     0 &     0 &     0 &    0 \\
 80$^\circ$ &     0 &     0 &     0 &     0 &     0 &     0 &     0 &    0 \\
 90$^\circ$ &     0 &     0 &     0 &     0 &     0 &     0 &     0 &    0 \\
\enddata
\tablecomments{Shown here is the number of comets ejected from the system for
pairs of initial conditions $(e_0,i_0)$.  The results of Fig.~\ref{Fig5} are
put in numbers here.  For low initial eccentricities, all comets stayed in
the system for the whole integration time of 1~Myr, whereas for high initial
eccentricities, i.e., $e_0 > 0.920$, comets are more likely to be ejected.
The number of ejected comets increases with increasing initial eccentricity.}
\label{Table3}
\end{deluxetable}

%+++++++++++++++++++++++++++++++++++++++++++++++++++++++++++++++++++++++

\clearpage

%%% *** Tab.4
%%%%%%%%%%%%%%%%%%%%%%%%%%%%%%%%%%%%%%%%%%%%%%%%%%%%%%%%%%%%%%%%%
\begin{deluxetable}{rrrrrrrrr}
\tablecaption{Results for Comets in Prograde Orbits, Part~2}
\tablewidth{0pt}
\tablehead{
Incl.  & \multicolumn{8}{c}{Eccentricity}  \\
\hline
\noalign{\smallskip}
...         & 0.945 & 0.950 & 0.955 & 0.960 & 0.965 & 0.970 & 0.975 & 0.980
}
\startdata
  0$^\circ$ &   542 &  1116 &  1919 &  2022 &  2530 &  4555 &  4099 &  4947 \\
 10$^\circ$ &   450 &  1005 &  1686 &  2500 &  3234 &  3360 &  3243 &  2908 \\
 20$^\circ$ &   337 &   630 &   947 &  1523 &  2068 &  1783 &  1535 &  1786 \\
 30$^\circ$ &   219 &   531 &   797 &  1196 &  1391 &  1428 &  1399 &  1544 \\
 40$^\circ$ &   191 &   385 &   548 &  1060 &  1239 &  1332 &  1136 &  1492 \\
 50$^\circ$ &     2 &     5 &    30 &   106 &   184 &   171 &   116 &   118 \\
 60$^\circ$ &     2 &     5 &    24 &    95 &   128 &   134 &    75 &   119 \\
 70$^\circ$ &     0 &     5 &    12 &    64 &   122 &   102 &    68 &    93 \\
 80$^\circ$ &    32 &    80 &   223 &   468 &   756 &   597 &   607 &   780 \\
 90$^\circ$ &     0 &     1 &     8 &    62 &    75 &    70 &    48 &    64 \\
\enddata
\tablecomments{
Continuation of Table~\ref{Table3}. For the high inclinations shown here,
the number of ejected comets strongly increases.}
\label{Table4}
\end{deluxetable}

%+++++++++++++++++++++++++++++++++++++++++++++++++++++++++++++++++++++++

\clearpage

%%% *** Tab.5
%%%%%%%%%%%%%%%%%%%%%%%%%%%%%%%%%%%%%%%%%%%%%%%%%%%%%%%%%%%%%%%%%
\begin{deluxetable}{rrrrrrrrr}
\tablecaption{Results for Comets in Retrograde Orbits}
\tablewidth{0pt}
\tablehead{
Incl.  & \multicolumn{8}{c}{Eccentricity}  \\
\hline
\noalign{\smallskip}
...           & 0.945 & 0.950 & 0.955 & 0.960 & 0.965 & 0.970 & 0.975 & 0.980
}
\startdata
  100$^\circ$ &     0 &     0 &     4 &    31 &    58 &    71 &    33 &    57 \\
  110$^\circ$ &     0 &     0 &     1 &    25 &    41 &    57 &    32 &    39 \\
  120$^\circ$ &     0 &     0 &     0 &    23 &    46 &    40 &    33 &    49 \\
  130$^\circ$ &     0 &     0 &     0 &    25 &    40 &    41 &    22 &    49 \\
  140$^\circ$ &     0 &     0 &     0 &    15 &    28 &    30 &    21 &    57 \\
  150$^\circ$ &     0 &     0 &     0 &    20 &    36 &    27 &    26 &    35 \\
  160$^\circ$ &     0 &     0 &     0 &     9 &    30 &    25 &    28 &    40 \\
  170$^\circ$ &     0 &     0 &     0 &     9 &    24 &    44 &    48 &    86 \\
\enddata
 \tablecomments{Compare Table~\ref{Table3} and \ref{Table4}. For retrograde orbits
($i_0 > 90^{\circ}$) the picture is qualitatively almost the same.  Comets starting
with initially low eccentricities tend to be stable and remain in the system.
Therefore, the number of ejected comets as, e.g., for $e_0=0.945$ is zero for all
retrograde inclinations.  For increasing values of eccentricity, the number of
ejected comets increases for all inclinations.}
\label{Table5}
\end{deluxetable}

%+++++++++++++++++++++++++++++++++++++++++++++++++++++++++++++++++++++++

\end{document}